\documentclass[aps,prd,showpacs,twocolumn,
superscriptaddress]{revtex4-1}

\usepackage{graphicx}
\usepackage{subfigure}
\usepackage{dcolumn}

\usepackage{amsmath}
\usepackage{amssymb}
\usepackage{bm}
\usepackage{color}
\usepackage[normalem]{ulem}
\usepackage[dvipsnames]{xcolor}
\usepackage{hyperref}

\usepackage{xcolor}
\usepackage{soul}

\graphicspath{{Figures/}}

\hypersetup{
colorlinks=true, % false: boxed links; true: colored links
linkcolor=blue,  % color of internal links
citecolor=cyan,  % color of links to bibliography
}

\newcommand{\bea}{\begin{eqnarray}}
\newcommand{\eea}{\end{eqnarray}}
\newcommand{\be}{\begin{equation}}
\newcommand{\ee}{\end{equation}}

\begin{document}
\title{
Gravitational waveform model based on photon motion for spinning black holes
}

\author{Song Li}
\email{lisong20@mails.ucas.ac.cn}
\affiliation{Shanghai Astronomical Observatory, Shanghai, 200030, China}
\affiliation{School of Astronomy and Space Science, University of Chinese Academy of Sciences,
Beijing, 100049, China}

\author{Wen-Biao Han}
\email{Corresponding author: wbhan@shao.ac.cn}
\affiliation{Shanghai Astronomical Observatory, Shanghai, 200030, China}
\affiliation{School of Fundamental Physics and Mathematical Sciences, Hangzhou Institute for Advanced Study, UCAS, Hangzhou 310024, China}
\affiliation{School of Astronomy and Space Science, University of Chinese Academy of Sciences,
Beijing, 100049, China}
\affiliation{Shanghai Frontiers Science Center for  Gravitational Wave Detection, 800 Dongchuan Road, Shanghai 200240, China}
\date{\today}
\begin{abstract}
The waveforms from binary black hole mergers include inspiral, merger, and ringdown parts. Usually, the inspiral waveform can be obtained by calibrating from post-Newtonian approximation; The merger and ringdown ones can be gotten from the quasinormal modes with black hole perturbation theory. However, for more general black holes, the calculation of the quasinormal modes is not trivial. In this paper we use the photon sphere to get the quasinormal modes of spinning black holes. Then we connect the ringdown wave with the inspiral part to get full  waveforms and compare with the ones from numerical relativity. We find that they match with each other very well. In principle this method can be extended to some general compact objects.  As an example, the ringdown waveforms from parametrized axisymmetric black holes are obtained. We also use this method to get the ringdown signals of the accelerating final black hole; this is due to gravitational recoil during the merger. Even for the extreme cases, the acceleration due to the recoil cannot produce detectable effects.

\end{abstract}


\maketitle
%%%%%%%%%%%%%%%%%%%%%%%%%%%%%%%%%%%%%%%%%%%%%%%%%
\section{Introduction}
There will be more gravitational-wave events in the future, so it is essential to have sufficiently-accurate waveform templates to study different gravitational wave events. There are already some templates for different situations: compact binaries on eccentric orbits\cite{Ecc_a, Ecc_b, seobnre, seobnre_s}, nonspinning  binaries\cite{Non_s_a, Non_s_b}, precessing binaries \cite{zhang2022}, etc., and it would be helpful if there was a template for different situations. For binary black hole(BH) events, there are three stages in the coalescence of two BHs: the long inspiral phase, during which the two black holes orbit each other; the merger phase, where the two black holes collide to form a single more heavy black hole; and the ringdown phase, where the remnant black hole emits gravitational waves(GWs) while settling down to its resting state. The inspiraling part can be described by the post-Newtonian approximation. At the end of the merger (i.e., the ringdown), the horizon of the remnant formed. At this moment, the spacetime is similar to a stable black hole but still has a slight perturbation. Therefore the merger and ringdown waveforms can be calculated from black-hole perturbation theory. The well-known EOBNR and IMRPhenom waveform models rely on the above idea\cite{IMRPhe_a, IMRPhe_b, IMRPhe_c, IMRPhe_d, IMRPhe_e, IMRPhe_f, IMRPhe_g, EOB_a, EOB_b, EOB_c, EOB_d, EOB_e, EOBNR_a, EOBNR_b, EOBNR_c, EOBNR_d, EOBNR_e, EOBNR_f, EOBNR_g}. Finally, from the gravitational waves emitted during the ringdown phase, one can get the information about the mass and spin of the final remnant\cite{MS_a, MS_b, MS_c}.

There are many different methods to calculate the perturbation. The common method uses the Teukolsky equation\cite{Teukolsky72}. There is a simple discussion of this method in \cite{Chandrasekhar84}. Considering the Teukolsky equation with a purely ingoing boundary at the horizon and purely outgoing boundary at infinity, the gravitational waves are characterized by some complex frequencies called quasinormal modes (QNMs)\cite{QNMs1a, QNMs1b, QNMs1c, QNMs1d, QNMs1e, QNMs1f}: the complex frequencies $\omega = \omega_R-i\omega_I$, where $\omega_R$ represents the frequency of GWs and $\omega_I$ represents the decay rate of GWs. Since QNMs arise as the perturbation of BHs spacetimes,  different BHs will have different QNMs. Chandrasekhar \textit{et al.}\cite{Chandrasekhar75a, Chandrasekhar75b, Chandrasekhar84} found that Schwarzschild BHs have two types of perturbations (scalar-type, and vector-type) characterized by the same QNMs spectrum. Some literature \cite{QNMs_Scha, QNMs_Schb, QNMs_Schc} gave the approximate analytic expressions of the QNMs for Schwarzschild BHs. The Kerr BHs have an additional parameter, the spin. Therefore the QNMs spectrum\cite{QNMs_Kerra, QNMs_Kerrb, QNMs_Kerrc, QNMs_Kerrd} is more complex than the Schwarzschild ones. The QNMs for a charged black hole, or for a black hole in an anti-de Sitter spacetime has also been studied in literature \cite{QNMs_chargeb, QNMs_chargec, QNMs_charged, QNMs_chargee, AdS1, AdS2, AdS3, AdS4, Review1, Review2, Review3}.

The QNMs are characterized by a small number of parameters: $s$, $l$, $m$, $n$. The field's spin is labeled as $s$, with $s=0$ for the scalar field, $s=\pm1$ for the electromagnetic fields, and $s=\pm2$ for the gravitational fields. $l$ and $m$ are harmonic indices($|m| \leq l$). $n$ is the overtone index that sorts the QNMs in order of decreasing damping timescales, so $n=0$ represents the least-damped mode(longest-lived mode or fundamental mode). Some works\cite{Higer_Overtonesa, Higer_Overtonesb} have found that using higher overtones can obtain a better description of the ringdown signal. Arnab\cite{Mirror_modes} proposed a negative-frequency (counterrotating) modes called mirror modes which are different from the positive-frequency (counterrotating) modes(e.g., higher overtones modes).

However, for non-GR(general relativity) black holes, direct calculation of QNMs is usually tricky. In \cite{QNMs_LR1, QNMs_chargea, Cardoso_09, Ikeda_21}, the authors found that the QNMs of the BHs with stationary, spherically symmetric, and asymptotically flat line elements are determined by the circular null geodesics, this result is based on the geodesic stability and Lyapunov exponents, on further studies, they found that for the axisymmetric cases, equatorial geodesics can account for the $l=m$ modes. Therefore, McWilliams\cite{BOB_PRL} proposed a new waveform model for the GR black holes called the backwards one-body(BOB) method, which does not include any phenomenological degrees of freedom. 

The study of photon motion has continued for decades: for examples, the ray-tracing codes based on photon motion \cite{PM_70s_a, PM_70s_b}, accretion disk\cite{PM_AD_a, PM_AD_b, PM_AD_c}, quasiperiodic oscillations\cite{PM_QPO_a}, and the first picture of the black hole(M87*)\cite{EHT_Fi, EHT_F, EHT_E}, which gives us a new channel to study black holes and general relativity. Some authors used the photon motion to study the BH shadow with different metrics: Kerr metric \cite{Chandrasekhar84}, Kerr-Newman metric\cite{shadow_b}, and some other rotating regular black holes\cite{shadow_c}. Some works have used the shadow of M87* to test the alternative theories of gravity: the superspinar\cite{M87_Sh_a}, the conformal massive gravity\cite{M87_Sh_b}, the Gauss-Bonnet gravity\cite{Shao20a}, Einstein-Maxwell-dilaton theory\cite{Martin22}, and symmergent gravity\cite{Irfan21}. In this paper, we use the photon motion to calculate the ringdown waveform for the objects in GR, such as the Kerr black hole and accelerating black hole, we also study the case with a general parameterized axisymmetric Konoplya, Rezzolla, and Zhidenko (KRZ) metric, this metric can be used to describe the nonstandard compact objects in GR and black holes in other gravity theories.

This paper is organized as follows: In Sec.~\ref{PM}, the photon motion under the Kerr metric is computed in two cases: the equatorial and arbitrary planes. In Sec.~\ref{Photon_QNMs}, we introduce how to use the photon motion (especially the photon sphere) to get the values(real part and imaginary part) of QNMs. In Sec.~\ref{Compare}, we compare the waveforms obtained from the light ring and NR (numerical relativity)/SEOBNR(spinning effective one body-numerical relativity) model. In Sec.~\ref{KRZ}, the ringdown waveform of a generally axisymmetric black hole is obtained. In Sec. \ref{ABH}, we study the effect of acceleration due to recoil on the gravitational waves. We conclude our results in Sec.~\ref{Conclusion}. We have fixed units such that $G = c = 1$.

\section{Photon motion}\label{PM}
In this section, we will review photon motion under the Kerr metric, and the photon motion equation will be used in the next section. The spacetime of the Kerr metric can be described by the following equation(we choose the Boyer-Linquist coordinates):
\begin{equation}
\begin{aligned}
d s^{2}=&-\left(1-\frac{2 M r}{\Sigma}\right) d t^{2}-\frac{4 M a r \sin ^{2} \theta}{\Sigma} d \phi d t+\frac{\Sigma}{\Delta} d r^{2} \\
&+\Sigma d \theta^{2}+\left(r^{2}+a^{2}+\frac{2 M r a^{2} \sin ^{2} \theta}{\Sigma}\right) \sin ^{2} \theta d \phi^{2},
\end{aligned}
\end{equation}
where
\begin{equation}
a \equiv J / M, \quad \Sigma \equiv r^{2}+a^{2} \cos ^{2} \theta, \quad \Delta \equiv r^{2}-2 M r+a^{2} .
\end{equation}
$M$, $J$, and $a$ are the mass, angular momentum, and angular momentum per unit mass of the black hole.

The Kerr metric is stationary and axisymmetric, and it is independent of $t$ and $\phi$ coordinates, which leads to the existence of timelike and spacelike Killing vectors. Consequently, there are two conserved quantities: the energy $E$ and the $z$-component $L_z$ of the angular momentum, which can be expressed as
\begin{equation}\label{E_Lz}
\begin{aligned}
-E &=g_{t t} \dot{t}+g_{t \phi} \dot{\phi}\,, \\
L_{z} &=g_{\phi t} \dot{t}+g_{\phi \phi} \dot{\phi}\,,
\end{aligned}
\end{equation}
where the overhead dot represents the derivative of the affine parameter; for photons, we can get the normalization condition of the four velocity:
\begin{equation}
u^{\alpha}u_{\alpha}=0\,,
\end{equation}
where $u^{\alpha}= (\dot{t}, \dot{r}, \dot{\theta}, \dot{\phi})$.

To study the motion of photons, we should use the geodesic equations, where we have for null geodesics the following expression:
\begin{equation}\label{Geodesic}
\frac{\mathrm{d}^{2} x^{\mu}}{\mathrm{d} \lambda^{2}}+\Gamma_{v \tau}^{\mu} \frac{\mathrm{d} x^{v}}{\mathrm{~d} \lambda} \frac{\mathrm{d} x^{\tau}}{\mathrm{d} \lambda}=0\,,
\end{equation}
where $\lambda$ is the affine parameter and $\Gamma_{v \tau}^{\mu}$ are the Christoffel Symbols defined as:
\begin{equation}
\Gamma_{v \tau}^{\mu}=\frac{1}{2} g^{\mu \sigma}\left(g_{\sigma v, \tau}+g_{\sigma \tau, v}-g_{v \tau, \sigma}\right)\,.
\end{equation}
We get the lightlike geodesic equation in separated two-order forms with Eq.~(\ref{Geodesic}). Though these two-order equations are not very difficult to solve, there is another way to get the lightlike geodesic equation in separated first-order forms. We have these conserved quantities: the energy $E$, the $z$ component $L_z$ of the angular momentum, and the four velocity $u^{\alpha}u_{\alpha}=0$, so we can use these conserved quantities to get the first-order equations. Since there are four directions for the photon motion, we must use another conserved quantity, which is the Carter constant $\mathcal{Q}$. Through these four conserved quantities: $E$, $L_z$, $u^{\alpha}u_{\alpha}=0$, and $\mathcal{Q}$, we can get the lightlike geodesic equation in the separated first-order form:
\begin{equation}\label{First_order}
\begin{aligned}
\Delta \Sigma \dot{t}&=\left[\left(r^{2}+a^{2}\right)^{2}-\Delta a^{2} \sin ^{2} \theta\right] E-2 M r a L_{z}\,, \\
\Sigma^{2} \dot{r}^{2}&=E^{2} r^{4}+\left(a^{2} E^{2}-L_{z}^{2}-\mathcal{Q}\right) r^{2}\\
&+2 M\left[\left(a E-L_{z}\right)^{2}+\mathcal{Q}\right] r-a^{2} \mathcal{Q}=R(r)\,, \\
\Sigma^{2} \dot{\theta}^{2}&=\mathcal{Q}-\left(\frac{L_{z}^{2}}{\sin ^{2} \theta}-E^{2} a^{2}\right) \cos ^{2} \theta\,, \\
\Delta \Sigma \dot{\phi}&=2 M r a E+(\Sigma-2 M r) \frac{L_{z}}{\sin ^{2} \theta}\,.
\end{aligned}
\end{equation}

Through Eq.~(\ref{First_order}), we can easily study the motion of photons. In the following sections, we will use Eq.~(\ref{First_order}) rather than Eq.~(\ref{Geodesic}).

%%%%%%%%%%%%%%%%%%%%%%%
\subsection{Equatorial plane}
If the photon is restricted on the equatorial plane, for the light ring, i.e., the circular orbit of the photon, $\dot{r}$ and $\dot{\theta}$ will both be equal to zero. As a result, Eq.~(\ref{First_order}) only retains two terms: $\dot{t}$ and $\dot{\phi}$. As we know, there is only one circular orbit (and unstable) for the Schwarzschild black hole and two circular orbits for the Kerr black hole: corotating and counterrotating light rings. We can get the properties (such as frequency) of the orbits through Eq. (\ref{First_order}) easily.

%%%%%%%%%%%%%%%%%%%%%%%
\subsection{Nonequatorial plane}
When considering the three-dimension (3D) motion, Eq.~(\ref{First_order}) should be calculated to get the photon orbits. The radius does not vary(i.e., $\dot{r}=0$) when considering the photon circular orbits. Then one can get $R(r)=\frac{dR(r)}{dr}=0$. For convenience, we use two new parameters($\tilde{L}$, $\tilde{Q}$) to replace the parameters ($E$, $L_z$, $\mathcal{Q}$), where $\tilde{L} \equiv L_z/E$ and $\tilde{Q} \equiv \mathcal{Q}/E$. Through this condition: $R(r)=\frac{dR(r)}{dr}=0$, we can get the expression of $\tilde{L}$ and $\tilde{Q}$:
\begin{equation}
\tilde{L} = -\frac{r^{3}-3 M r^{2}+a^{2} r+a^{2} M}{a(r-M)}\,,
\end{equation}

\begin{equation}
\tilde{Q} = -\frac{r^{3}\left(r^{3}-6 M r^{2}+9 M^{2} r-4 a^{2} M\right)}{a^{2}(r-M)^{2}}\,.
\end{equation}

In this case, the photons oscillate around the equatorial plane. The maximum angle can be solved through the following equation\cite{PO_a}:
\begin{equation}
\begin{aligned}
u_{0}^{2}&=\cos ^{2} \theta_{\max } \\
&=\frac{1}{2 a^{2}}\left[\left(a^{2}-\tilde{Q}-\tilde{L}^{2}\right)+\sqrt{\left(a^{2}-\tilde{Q}-\tilde{L}^{2}\right)^{2}+4 a^{2} \tilde{Q}}\right]\,.
\end{aligned}
\end{equation}

Using Eq.~(\ref{First_order}), considering there is a square on the $\dot{\theta}$ (i.e., $\dot{\theta}^2$) so we use a new parameter $u$ to replace $\theta$.
\begin{equation}
u = \cos \theta\,,
\end{equation}
then we can rewrite the equation of $\theta$ as
\begin{equation}
\Sigma^{2} \dot{u}^{2}=a^{2}\left(u_{0}^{2}-u^{2}\right)\left(u^{2}-u_{1}^{2}\right) \,,
\end{equation}
where 
\begin{equation}
u_{1}^{2}=\frac{1}{2 a^{2}}\left[\left(a^{2}-\tilde{Q}-\tilde{L}^{2}\right)-\sqrt{\left(a^{2}-\tilde{Q}-\tilde{L}^{2}\right)^{2}+4 a^{2} \tilde{Q}}\right] \,.
\end{equation}
By rewriting $u = u_0\sin \chi(\lambda)$, where $\chi$ is a monotonic function, we get the following equation:
\begin{equation}\label{theta_new}
\dot{\chi}=\pm \frac{a}{\Sigma} \sqrt{u^{2}-u_{1}^{2}}=\pm \frac{a}{r^{2}+a^{2} u_{0}^{2} \sin ^{2} \chi} \sqrt{u_{0}^{2} \sin ^{2} \chi-u_{1}^{2}} \,.
\end{equation}
Now we use Eq.~(\ref{theta_new}) to replace the $\theta$ direction equation in Eq. (\ref{First_order}), then, combining $\dot{t}$ and $\dot{\phi}$ in Eq.~(\ref{First_order}), the circular orbits (light sphere) in 3D can be solved. A more detailed description of the calculations is presented in \cite{PO_a}.

%%%%%%%%%%%%%%%%%%%%%%%%%%%%%%%%%%%%%%%%%%%%%%%%%
\section{Getting QNMs from photon motions}\label{Photon_QNMs}
Photon motion (``null orbit") is essential for astrophysical and theoretical study. For example, the optical appearance to the observer of a collapsed star is related to the properties of photon motion. In addition, photon motion is also related to the characteristic modes(QNMs) of a black hole in the eikonal limit\cite{Goebel72, Ferrari84, Cardoso_09, Ikeda_21}. These modes can be approximated related with the terms of photons moving in an unstable circular orbit (also known as  ``light ring" or ``photon sphere"). The decay rate ($\omega_I$) in QNMs connects with the Lyapunov exponent of light rays (the neighboring two photons on the light ring will separate exponentially due to the unstable orbit). As the same as $\omega_I$, the parameter $\omega_R$ can be represented by the energy of the photon sphere. Therefore the gravitational-wave emissions at the ringdown period could be described by the properties of the null geodesics on unstable circular orbits at the black hole's light ring. So, in this section, we will show how to connect the photon motion and the QNMs. The complex frequencies in QNMs is $\omega= \omega_R-i\omega_I$. We will introduce how to get $\omega_R$ and $\omega_I$ through the photon motion. There are three parameters in QMNs: the multipolar indices $l$ and $m$ and the overtone index $n$(the parameter $s$ is a constant in gravitational fields). This section considers the case where the overtone index is the fundamental mode $n=0$.

%%%%%%%%%%%%%%%%%%%%%%%
\subsection{Determination of $\omega_R$}
%%%%%%%%%%%%%%%%%%%%%%%
We get the value of $\omega_R$ by following Ref. \cite{QNMs_LR_a}, which considers the photon in 3D motion. The real part of the QNMs, $\omega_R$, can be decomposed along $\theta$ and $\phi$:
\begin{equation} \label{omega_R_equ}
\omega_R = L\Omega_{\theta}(m/L)+m\Omega_{\rm prec}(m/L)\,,
\end{equation}
%%%%%%%%%%%%%%%
where $\Omega_{\theta}$ is the frequency of polar motion, the frequency at which the photon oscillates below and above the equatorial plane, with a period given by $T_{\theta} = 2\pi/\Omega_{\theta}$.

The particle also moves in the azimuthal ($\phi$) direction in the period $T_{\theta}$. Usually, when the polar motion finishes a period, $\phi$ does not scan through $2\pi$ for a corotating orbit($m>0$) or $-2\pi$ for a counterrotating orbit($m<0$) due to the relativistic precession. The difference between the $\Delta \phi$ and $\pm2\pi$(its precession-free value) is denoted as the ``precession angle'':
%%%%%%%%%%%%%%%
\begin{equation}
\Delta\phi_{\rm prec} = \Delta\phi - 2\pi {\rm sgn} (m)\,,
\end{equation}
%%%%%%%%%%%%%%%
\begin{equation}
\Omega_{\rm prec} = \Delta\phi_{\rm prec}/T_{\theta}\,,
\end{equation}
%%%%%%%%%%%%%%%
\begin{equation}
L = l+1/2\,,
\end{equation}
%%%%%%%%%%%%%%%
the value of $m$ is a constant that can be freely chosen(in this paper, we choose $m=2$), and the value of $l$ can be obtained through the following content.

For orbits between the equatorial orbits and polar ones, we can use the following equations:
\begin{equation}\label{V_r_T}
\begin{aligned}
\Omega_{R} &=\frac{\mu a}{r_{0}^{2}+a^{2}} \pm \frac{\sqrt{\Delta\left(r_{0}\right)}}{r_{0}^{2}+a^{2}} \beta\left(a \Omega_{R}\right)\,, \\
0 &=\frac{\partial}{\partial r}\left[\frac{\Omega_{R}\left(r^{2}+a^{2}\right)-\mu a}{\sqrt{\Delta(r)}}\right]_{r=r_{0}}\,,
\end{aligned}
\end{equation}
where $\Omega_R = \frac{\omega_R}{L}$, $\mu = \frac{m}{L}$.

Equation~(\ref{V_r_T}) is gotten through the following equation:
\begin{equation}
V^{r}\left(r, \omega_{R}\right)=\left.\frac{\partial V^{r}}{\partial r}\right|_{\left(r, \omega_{R}\right)}=0\,,
\end{equation}
where $V^{r}$ is the potential in the radial Teukolsky equation.

\begin{equation}
\begin{aligned}
\beta(z) &=\sqrt{\alpha(z)+z^{2}-2 \mu z} \\
& \approx \sqrt{1+\frac{z^{2}}{2}-2 \mu z+\frac{\mu^{2} z^{2}}{2}}\,,
\end{aligned}
\end{equation}
for convenience, we can use another method to get the value of $l$:
\begin{equation}\label{solve_l}
\begin{aligned}
\Omega_{R} &=\frac{\mu a}{r_{0}^{2}+a^{2}} \pm \frac{\sqrt{\Delta\left(r_{0}\right)}}{r_{0}^{2}+a^{2}} \beta\left(a \Omega_{R}\right), \\
 &=\frac{\left(M-r_{0}\right) \mu a}{\left(r_{0}-3 M\right) r_{0}^{2}+\left(r_{0}+M\right) a^{2}} \,.
\end{aligned}
\end{equation}
One can obtain $r_0$ and $\mu$ from the above equations. Therefore, since $\mu=m/L$, $m=2$, and $L=l+1/2$ we can get the value of $l$.

We do not choose a specific metric when computing the value of $\omega_R$. We only write down the photon motion equations, then, with these equations, we can get the value of $\omega_R$, this result is valid in the eikonal approximation, even for small values $l$. Thus, we can apply this method to some GR and non-GR black hole metrics (which we will show in the following sections).

\subsection{Determination of $\omega_I$}
%%%%%%%%%%%%%%%
The ringdown signal damping rate:
\begin{equation}\label{gamma_1}
\omega_{I}=\left(n+\frac{1}{2}\right) \frac{\sqrt{2 \mathcal{R}_{0}^{\prime \prime}} \Delta_{0}}{\left[\frac{\partial \mathcal{R}}{\partial \mathcal{E}}+\frac{\partial \mathcal{R}}{\partial \mathcal{Q}}\left(\frac{d \mathcal{Q}}{d \mathcal{E}}\right)\right]_{r_{0}}} \,,
\end{equation}
where
\begin{equation}\label{gamma_2}
\mathcal{R}(r)=\left[{E}\left(r^{2}+a^{2}\right)-L_{z} a\right]^{2}-\Delta\left[\left(L_{z}-a {E}\right)^{2}+\mathcal{Q}\right] \,,
\end{equation}
and
\begin{equation}\label{gamma_3}
\mathcal{Q} \approx L^{2}-m^{2}-\frac{a^{2} \omega_{R}^{2}}{2}\left[1-\frac{m^{2}}{L^{2}}\right] \,.
\end{equation}
Through Eqs.~(\ref{gamma_1}), (\ref{gamma_2}), and (\ref{gamma_3}) we can get the value of $\omega_I$ with different $r_0$. Applying the same method to obtain $\omega_R$, we can derive the expression of $\omega_I$ only by using the photon motion equations. As a result, this method can be applied to some other black holes.

\begin{figure}
\includegraphics[width=0.5 \textwidth]{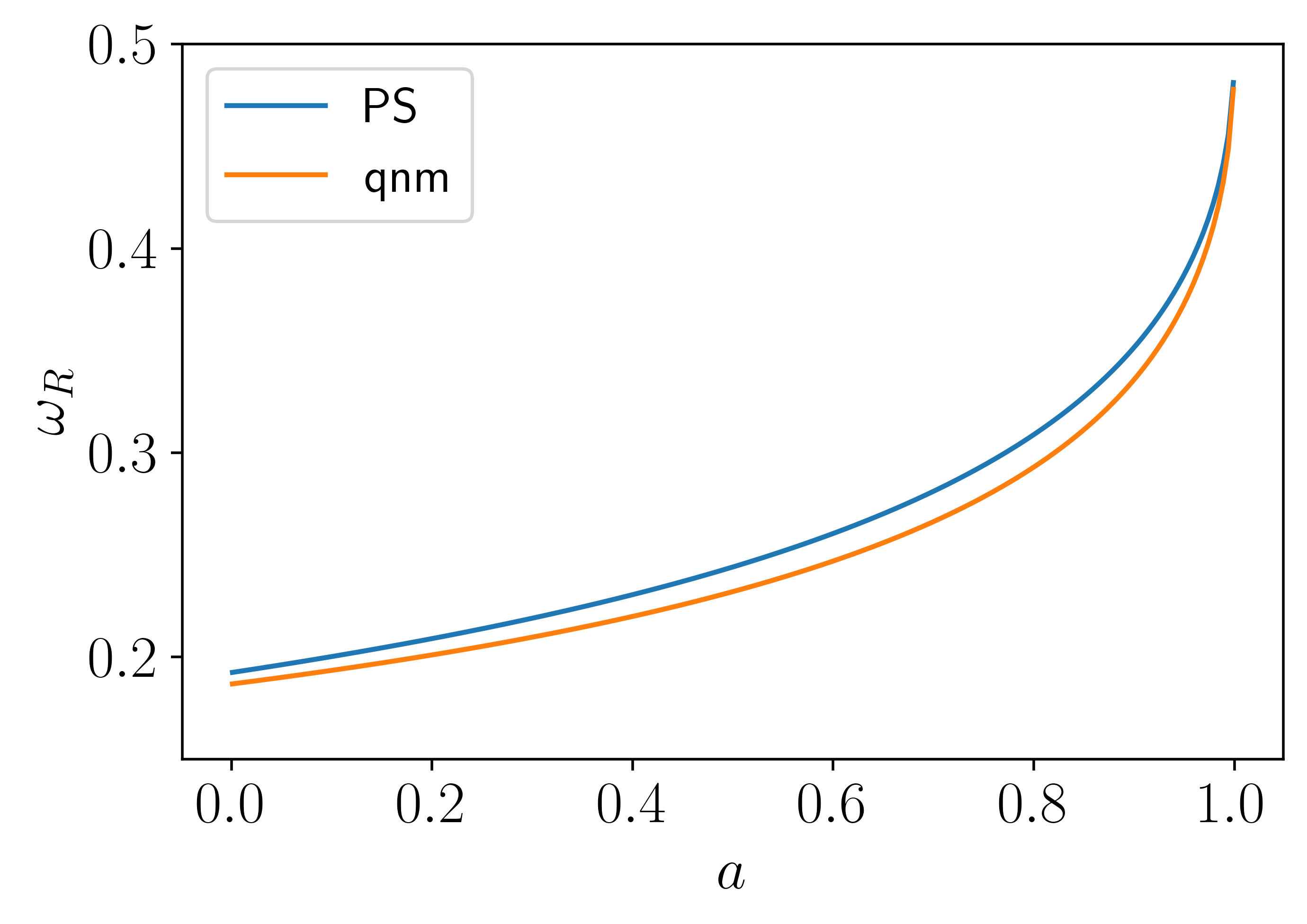}
\caption{Comparison of $\omega_R$ calculated from the photon sphere (PS) and black hole perturbation method.\label{Omega_R}}
\end{figure}

\begin{figure}
\includegraphics[width=0.5 \textwidth]{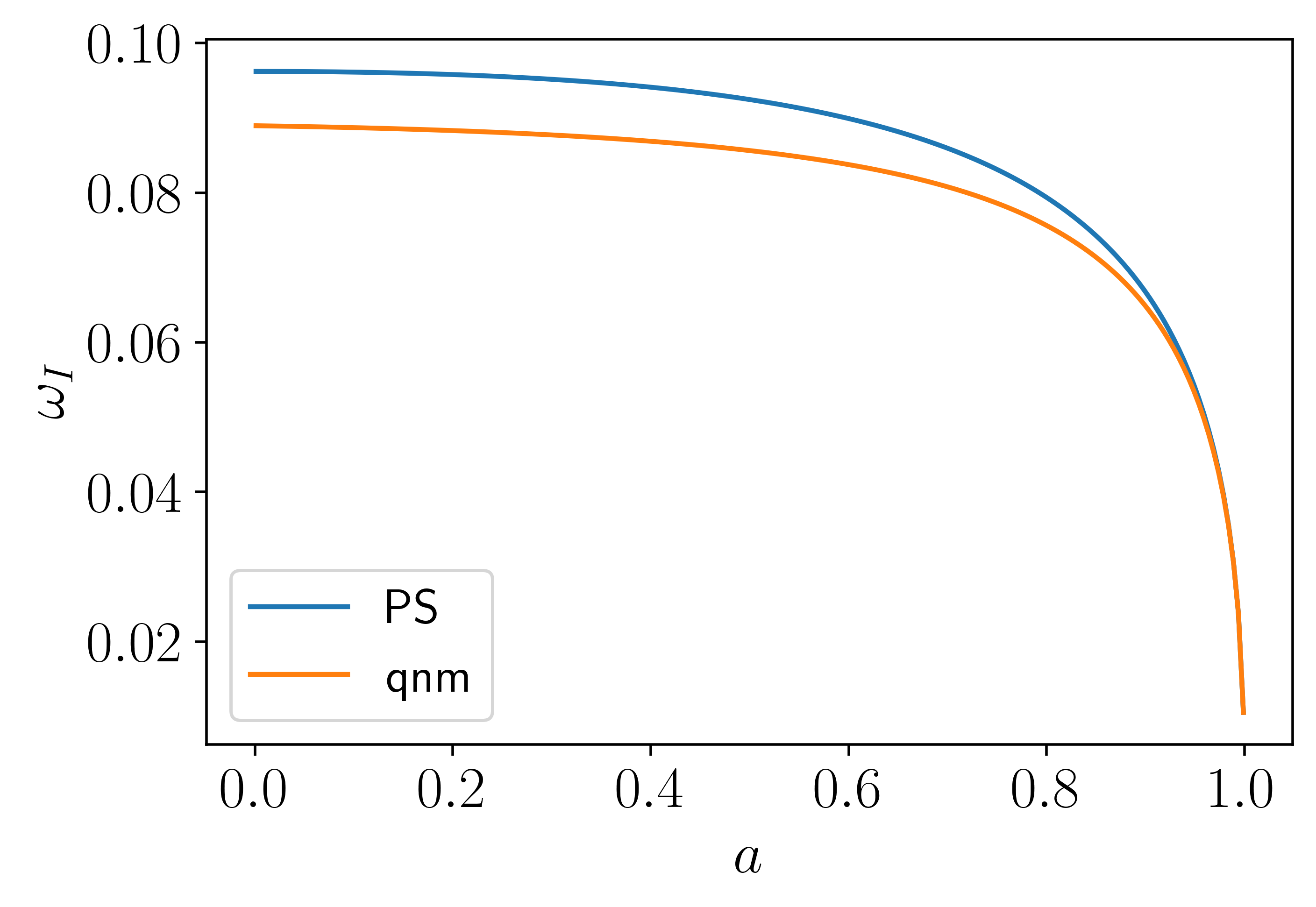}
\caption{Comparison of $\omega_I$ calculated from the PS and black hole perturbation method.\label{Omega_I}}
\end{figure}

The above calculations show the procedure to get the quantities $\omega_R$ and $\omega_I$ through the photon motion. To test the accuracy of the calculated values of $\omega_R$ and $\omega_I$ we obtained, we use the package ``qnm'' \cite{qnm}(qnm is an open-source PYTHON package for computing the Kerr quasinormal mode frequencies, angular separation constants, and spherical-spheroidal mixing coefficients).
As shown in Figs.~\ref{Omega_R}, \ref{Omega_I}, we observe that the $\omega_R$ and $\omega_I$ calculated from the photon motion have good accuracy at high spin. We suppose this could be because the photon sphere is closer to the horizon as the spin increases and will have a better reflection of the BHs. The maximum relative difference of $\omega_R$ is not larger than 5.6\%, and $\omega_I$ is not larger than 8.5\%. One thing that should be mentioned is that the parameter $\omega_R$ is the wave frequency, which is the same as the energy ${E}$. Therefore when calculating $\left(\frac{d \mathcal{Q}}{d {E}}\right)$ we can just calculate $\left(\frac{d \mathcal{Q}}{d \omega_R}\right)$ through Eq.~(\ref{gamma_3}).

%%%%%%%%%%%%%%%%%%%%%%%%%%%%%%%%%%%%%%%%%%%%%%%%%
\section{Gravitational wave and comparison with NR/SEOBNRv4}\label{Compare}
This section will use $\omega_R,~ \omega_I$ obtained from the photon sphere to construct the ringdown signals. Furthermore, we connect the PN inspiral waves with the ringdown one to derive the full waveforms, so we call our model PSI ($\Psi$: photon sphere + inspiral) waveforms. The PSI model is inspired by the BOB waveform model described in Ref. \cite{BOB_PRL}. This model calculates the waveforms for equal mass and the GR binary black holes by the light ring. Many waveform models include phenomenological degrees (e.g., phenomenological coefficients) to have good accuracy with NR. However, the BOB method does not include phenomenological degrees of freedom but has a good result compared with NR in the ringdown period. In the present paper, we will construct our $\Psi$ waveforms for the binary BH mergers with varied spins and mass ratios, then apply them to the parametrized (KRZ) black holes and accelerating black holes.

As shown in \cite{BOB_PRL, BOB_b}, the amplitude of the GW has the following form:
\begin{equation}
{{\left| {{h}_{lm}} \right|}^{2}}\sim \frac{d}{dt}\left( {{\Omega }_{lm}}^{2} \right)\,,
\end{equation}
where $\Omega_{lm}$ is the orbital frequency; through this equation, we can get the equation of the GW waveform:

\begin{equation}
{h}_{22}=X \operatorname{sech}\left[\gamma\left(t-t_{p}\right)\right] e^{-i \tilde{\Phi}_{22}(t)}\,,
\end{equation}
where $X$ is a constant, $\gamma$ is the Lyapunov exponent, $t_p$ is the time corresponding to the peak waveform amplitude, and $\Phi_{22}(t)$ is the phase.

We can also derive the phase equation:
\begin{equation}
\begin{aligned}
\tilde{\Phi}_{22}=& \int_{0}^{t} \Omega d t^{\prime}=\arctan _{+}+\operatorname{arctanh}_{+} \\
&-\arctan _{-}-\operatorname{arctanh}_{-}-\phi_0,
\end{aligned}
\end{equation}

where
\begin{equation}
\left\{\begin{array}{c}
\arctan _{\pm} \equiv \kappa_{\pm} \tau\left[\arctan \left(\frac{\Omega}{\kappa_{\pm}}\right)-\arctan \left(\frac{\Omega_{0}}{\kappa_{\pm}}\right)\right]\,, \\
\arctan \mathrm{h}_{\pm} \equiv \kappa_{\pm} \tau\left[\arctan h\left(\frac{\Omega}{\kappa_{\pm}}\right)-\arctan h\left(\frac{\Omega_{0}}{\kappa_{\pm}}\right)\right]
\end{array}\right.
\end{equation}
\begin{equation}
\kappa_{\pm} \equiv\left\{\Omega_{0}^{4} \pm k\left[1 \mp \tanh \left(\frac{t_{0}-t_{p}}{\tau}\right)\right]\right\}^{1 / 4}\,,
\end{equation}

\begin{equation}\label{attention1}
\Omega=\left\{\Omega_{0}^{4}+k\left[\tanh \left(\frac{t-t_{p}}{\tau}\right)-\tanh \left(\frac{t_{0}-t_{p}}{\tau}\right)\right]\right\}^{1 / 4}\,,
\end{equation}

\begin{equation}\label{attention2}
k=\left(\frac{\Omega_{\mathrm{QNM}}^{4}-\Omega_{0}^{4}}{1-\tanh \left[\left(t_{0}-t_{p}\right) / \tau\right]}\right)\,,
\end{equation}\\
where $\tau=\gamma^{-1}$, ${{\Omega }_{\operatorname{QNM}}}$=$\omega_{\operatorname{QNM}}$/m($\Omega_{\operatorname{QNM}}$ is just $\omega_R$), and $\phi_0$, $\Omega_0$, $t_0$ are the constants that can be freely chosen.

We should pay attention to Eqs.~(\ref{attention1}) and (\ref{attention2}) in the above equations, the presence of terms with an even power imposes an additional condition on $\Omega_0$. We want to find the minimum value of $\Omega_0$, so we can let the value inside Eq.~(\ref{attention1}) equal to zero. Then we obtain this function:
\begin{equation}\label{attention3}
\Omega_{0}^{4}=k\left[-\tanh \left(\frac{t-t_{p}}{\tau}\right)+\tanh \left(\frac{t_{0}-t_{p}}{\tau}\right)\right]\,.
\end{equation}
Substituting Eq.~(\ref{attention2}) into Eq.~(\ref{attention3}), we can get the solution of Eq.~(\ref{attention3})(we only consider the positive solution):
\begin{equation}\label{Omega_min}
{{\Omega }_{0}}^{4}=\frac{{{\Omega }_{\text{QNM}}}^{4}(\tanh [\frac{t-{{t}_{p}}}{\tau }]-\tanh [\frac{{{t}_{0}}-{{t}_{p}}}{\tau }])}{(-1+\tanh [\frac{{{t}_{0}}-{{t}_{p}}}{\tau }])(1-\frac{\tanh [\frac{t-{{t}_{p}}}{\tau }]}{1-\tanh [\frac{{{t}_{0}}-{{t}_{p}}}{\tau }]}+\frac{\tanh [\frac{{{t}_{0}}-{{t}_{p}}}{\tau }]}{1-\tanh [\frac{{{t}_{0}}-{{t}_{p}}}{\tau }]})}\,.
\end{equation}\\

\begin{figure}
\includegraphics[width=0.5 \textwidth]{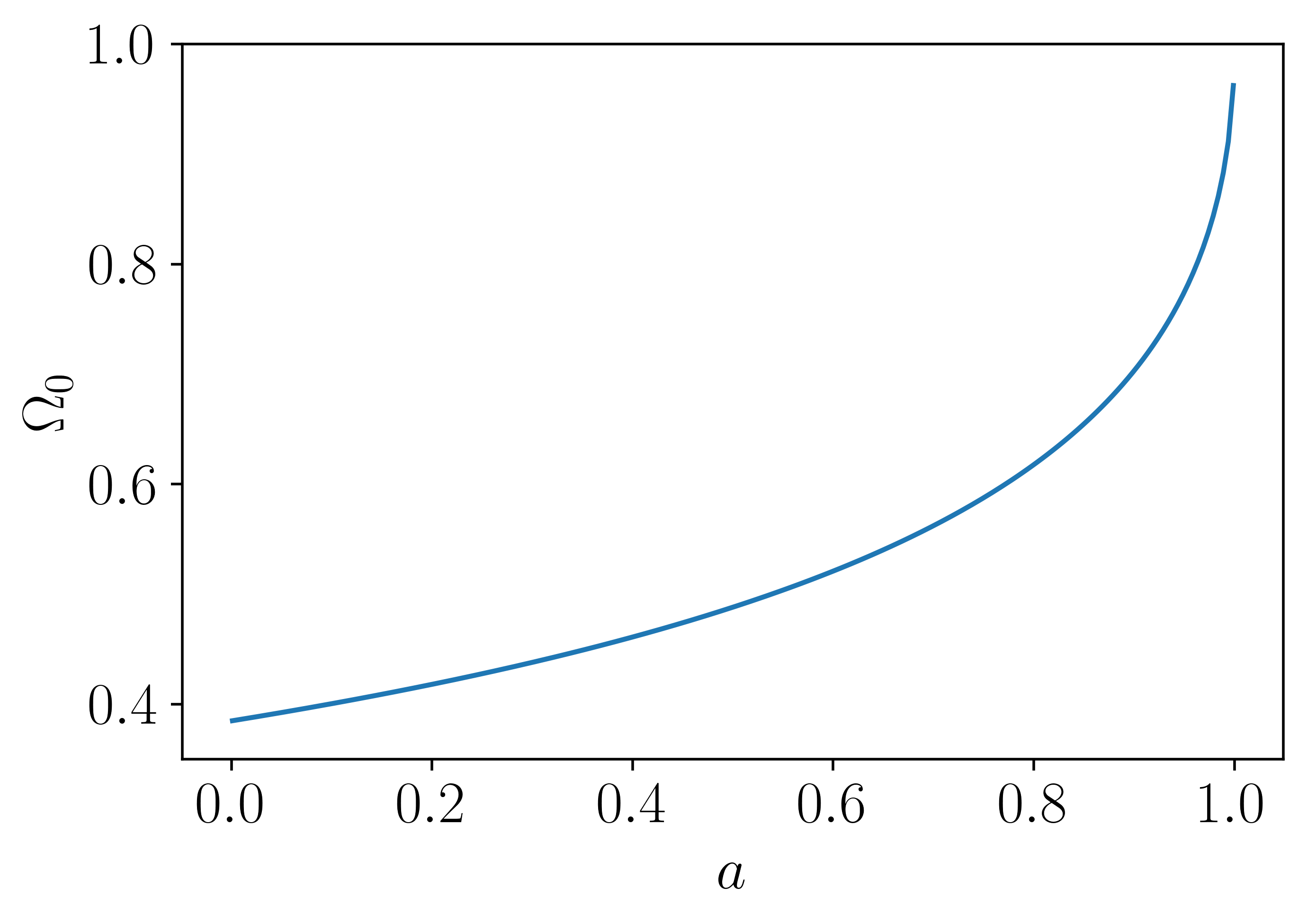}
\caption{The minimum allowable value of $\Omega_0$ in the BOB model with different spins under Kerr metric. \label{Omega_0_min}}
\end{figure}

With Eq.~(\ref{Omega_min}), we get the minimum value of $\Omega_0$. For convenience, we choose $t$ to be equal to $t_p$, so Eq.~(\ref{Omega_min}) can be simplified to this form:
\begin{equation}\label{Omega_min_simplify}
{{\Omega }_{0}}^{4}={{\Omega }_{\text{QNM}}}^{4}(\tanh [\frac{{{t}_{0}}-{{t}_{p}}}{\tau }])\,.
\end{equation}\\
Thus, we obtain the minimum value of $\Omega_0$ is $\Omega_{\operatorname{QNM}}$(i.e., the region of $\Omega_0$ is $\Omega_0 \textgreater \Omega_{\operatorname{QNM}}$). Figure \ref{Omega_0_min} shows the minimum value of  $\Omega_0$ with different spins in Kerr BHs. Because the value of  $\Omega_0$ relates to $\Omega_{\operatorname{QNM}}$, the minimum value of $\Omega_0$ is different for different metrics.

\begin{figure*}
\centering
\subfigure[$a$ = 0.10]{
\includegraphics[width=0.4 \textwidth]{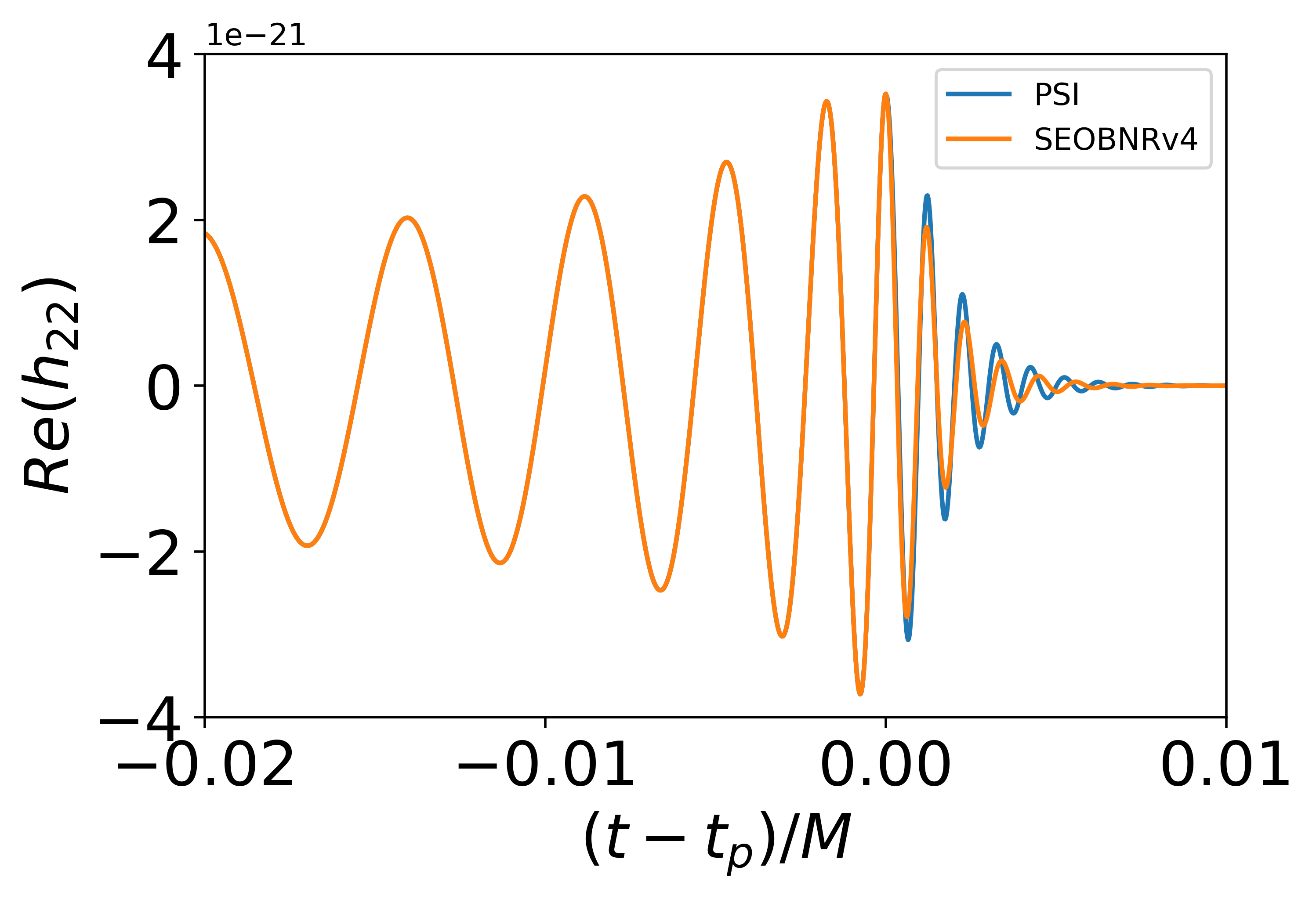}
}
\quad
\subfigure[$a$ = 0.20]{
\includegraphics[width=0.4 \textwidth]{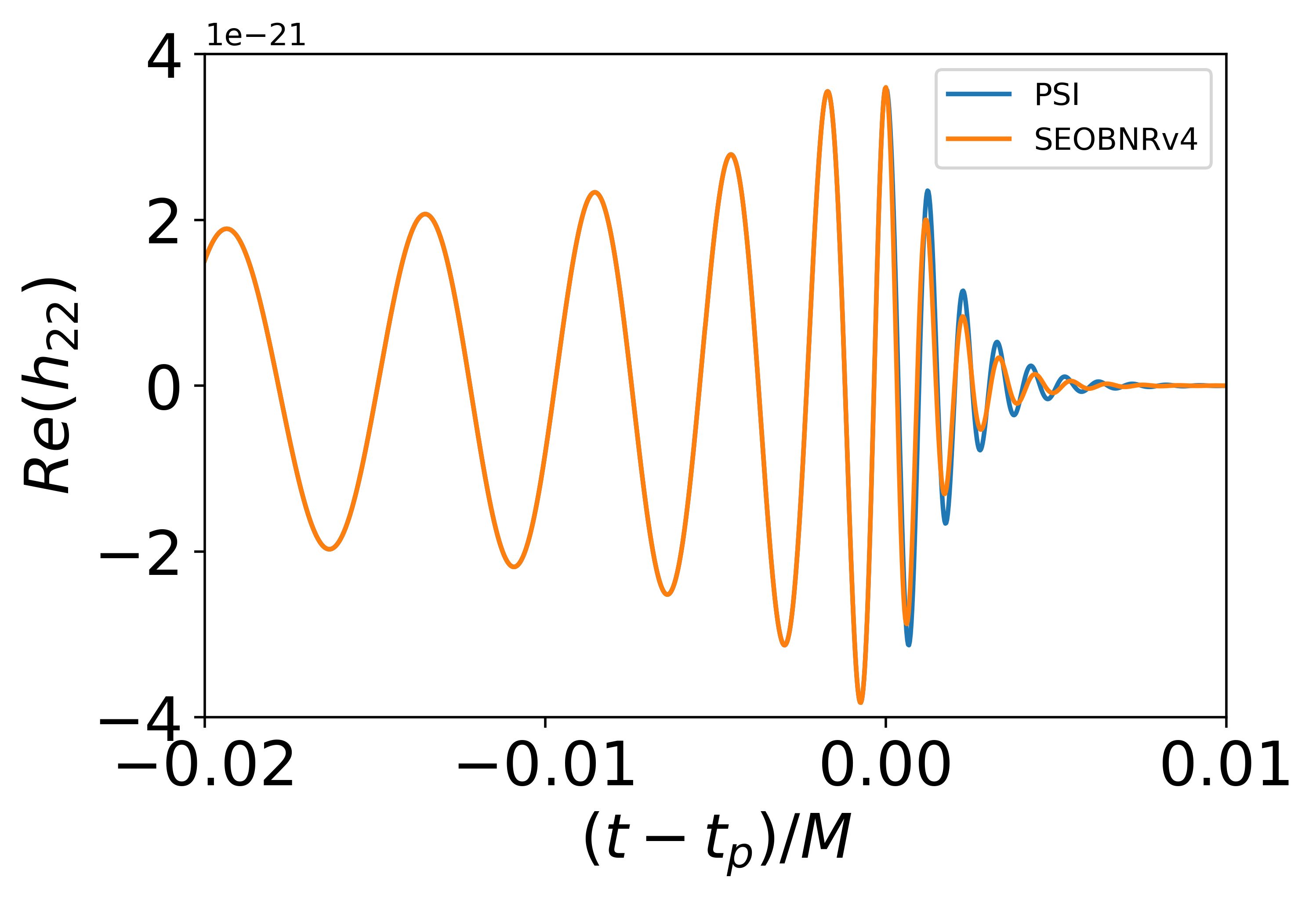}
}
\quad
\subfigure[$a$ = 0.30]{
\includegraphics[width=0.4 \textwidth]{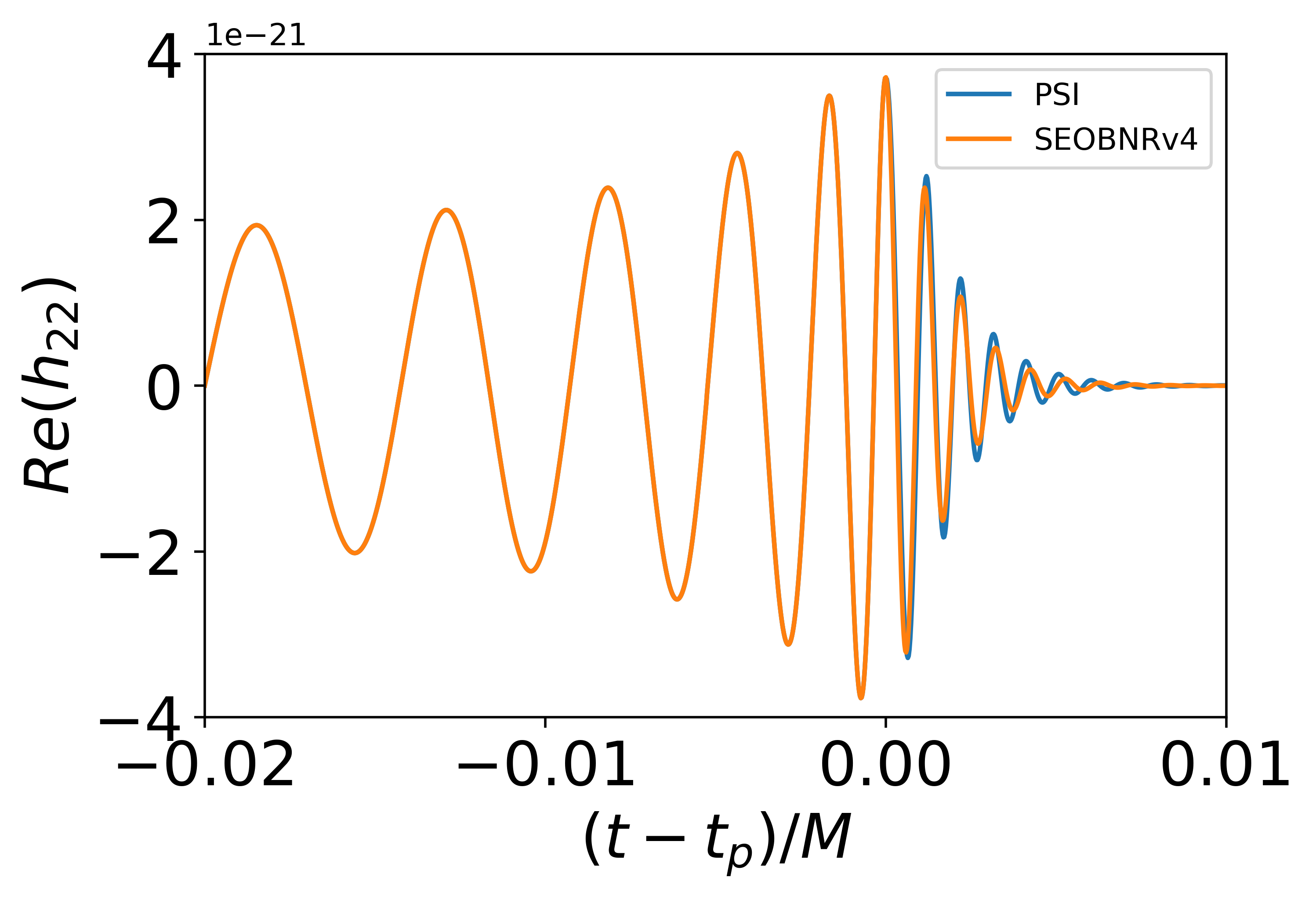}
}
\quad
\subfigure[$a$ = 0.40]{
\includegraphics[width=0.4 \textwidth]{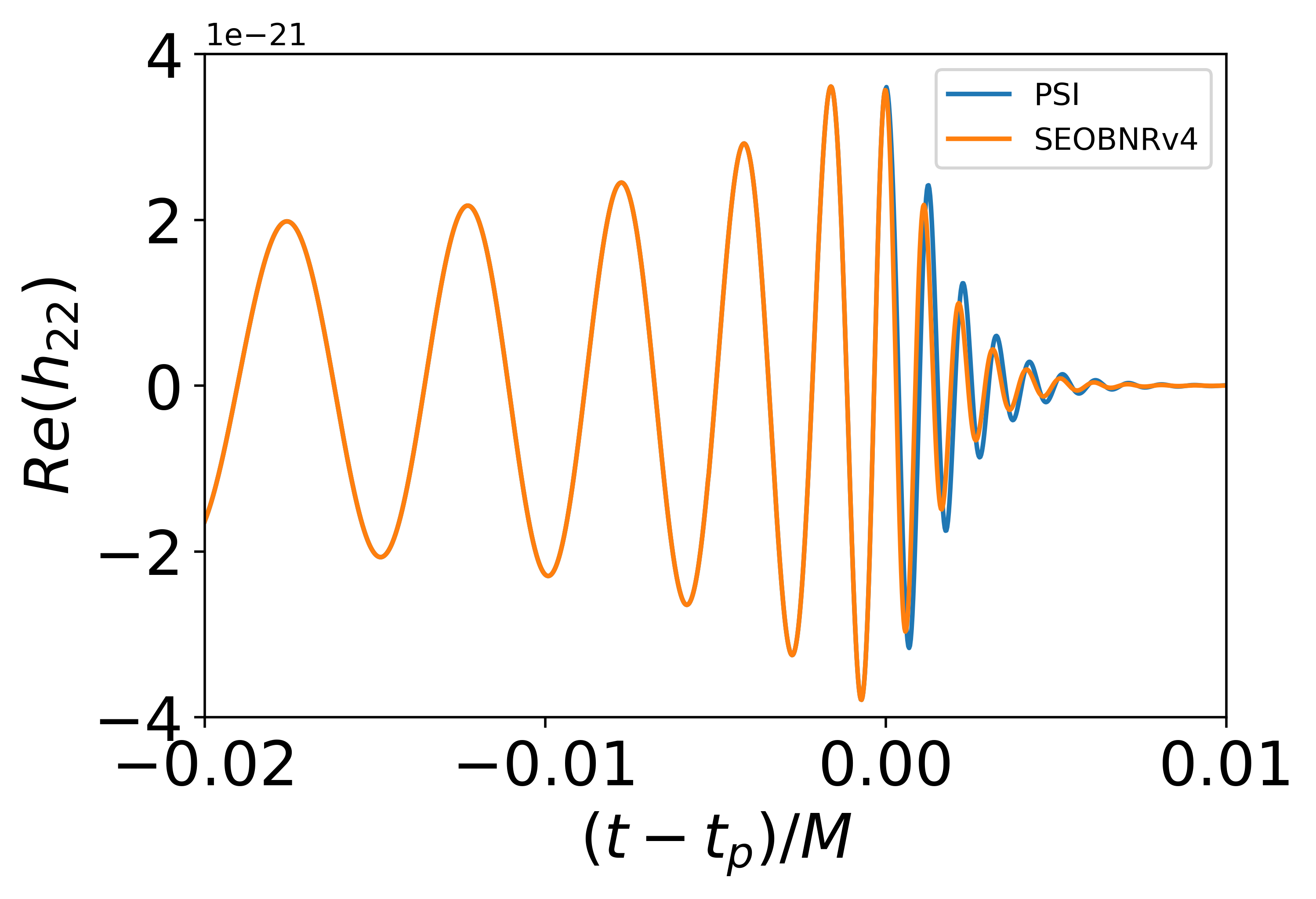}
}
\quad
\subfigure[$a$ = 0.50]{
\includegraphics[width=0.4 \textwidth]{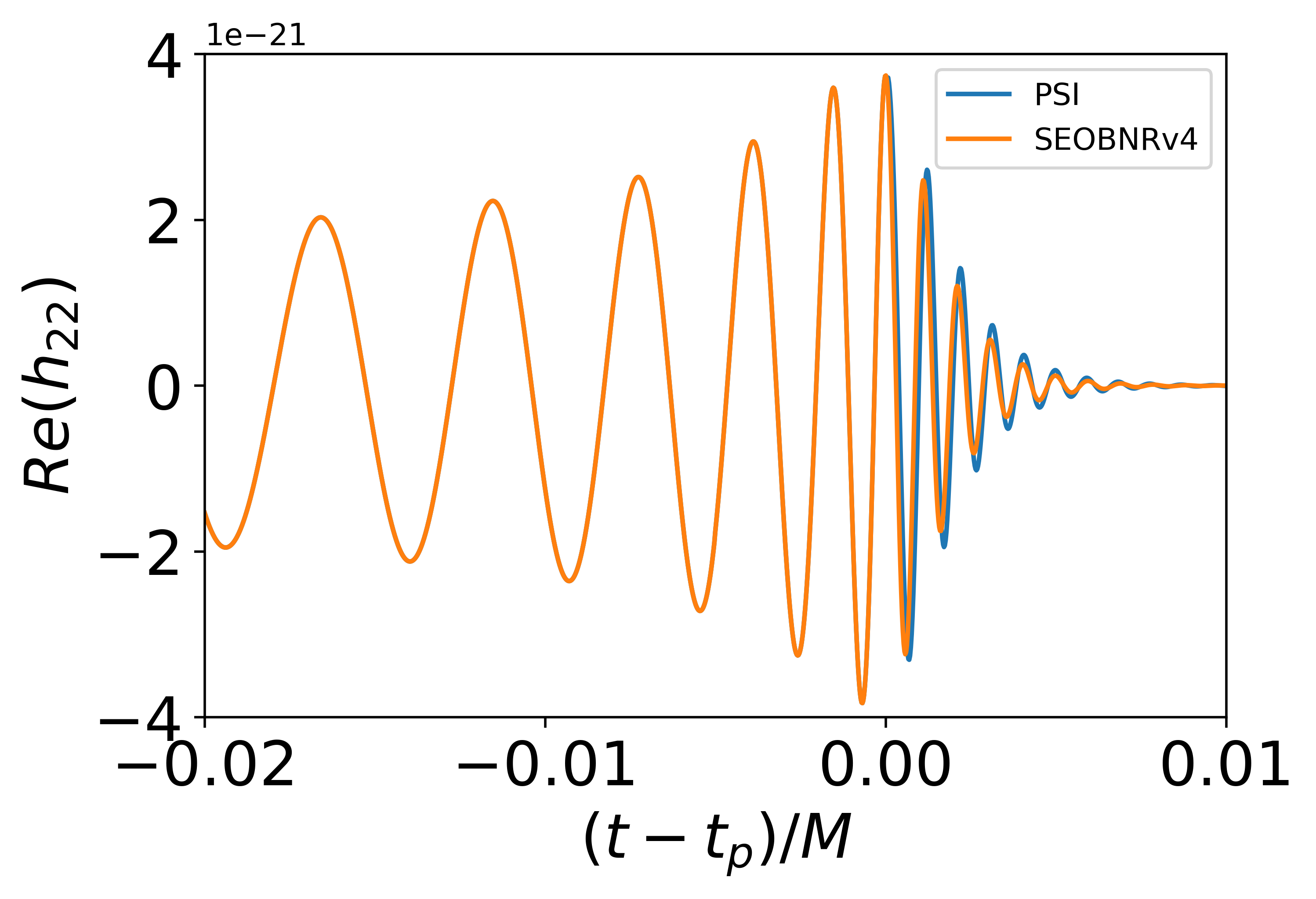}
}
\quad
\subfigure[$a$ = 0.60]{
\includegraphics[width=0.4 \textwidth]{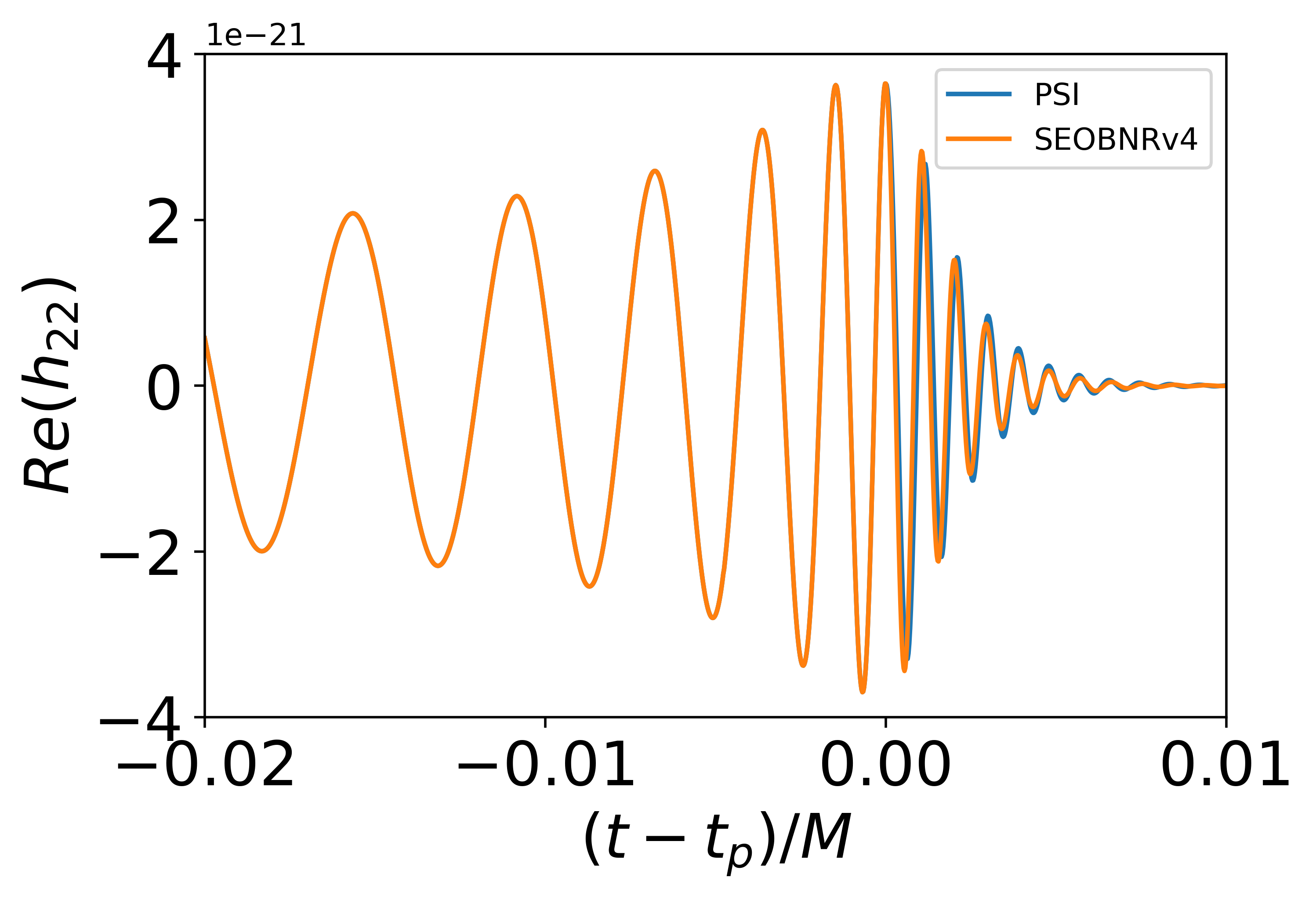}
}
\caption{Comparison of PSI ($\Psi$) and SEOBNRv4 waveforms with different spins. In the inspiral part, the two kind waveforms are totally overlapped due to the PSI model including the EOB inspiral waves.\label{compare1}}
\end{figure*}

Now we need to connect the waveforms from the photon sphere with the inspiral ones. This match can be done anywhere between the innermost stable circular orbit and the light ring. In the present work, we choose the peak of the waveform as the match point. Then we can get the optimal values for $\phi_0$, $\Omega_0$, and $t_0$. First, to validate our model, we compare the $\Psi$ waveforms with SEOBNRv4 ones. We plot the waveforms with different spins to compare with SEOBNRv4 and PSI in Fig.~\ref{compare1}. We use parameter overlap to make a more precise judgment on the PSI model's accuracy. The definition of the overlap is as follows:

\begin{equation}
\mathrm{F}= \left[\frac{\left\langle h_{1} \mid h_{2}\right\rangle}{\sqrt{\left\langle h_{1} \mid h_{1}\right\rangle\left\langle h_{2} \mid h_{2}\right\rangle}}\right],
\end{equation}

\begin{equation}
\left\langle h_{1}, h_{2}\right\rangle=4 \operatorname{Re} \int_{f_{\min }}^{f_{\max }} \frac{\tilde{h}_{1}(f) \tilde{h}_{2}^{*}(f)}{S_{n}(f)} d f,
\end{equation}

where $h_1$ is the waveform derived from the PSI model, $h_2$ is the compared waveform(e.g., SEOBNRv4, SXS), and $S_n(f)$ is the power spectral density of the detector noise, in this work we use the aLIGO's sensitivity curve\cite{Harry_2010}. And the definition of the match is
\begin{equation}
\mathrm{FF}=\max \left[\frac{\left\langle h_{1} \mid h_{2}\right\rangle}{\sqrt{\left\langle h_{1} \mid h_{1}\right\rangle\left\langle h_{2} \mid h_{2}\right\rangle}}\right],
\end{equation}
and the mismatch of two waveforms is defined as $1-\mathrm{FF}$.

From Fig.~\ref{compare1}, we find that the PSI model coincides with SEOBNRv4 one very well, especially at high spin. Even at low spin, the overlap is still larger than 98\%. Note that here we only compare the ringdown part. The overlap will be larger if we compare the complete waves because our PSI waveform includes the EOB inspiral signal. Therefore we conclude that the PSI waveforms have a good consistency with SEOBNRv4 ones for spinning binary black holes.

\begin{figure*}
\centering
\includegraphics[width=0.4 \textwidth]{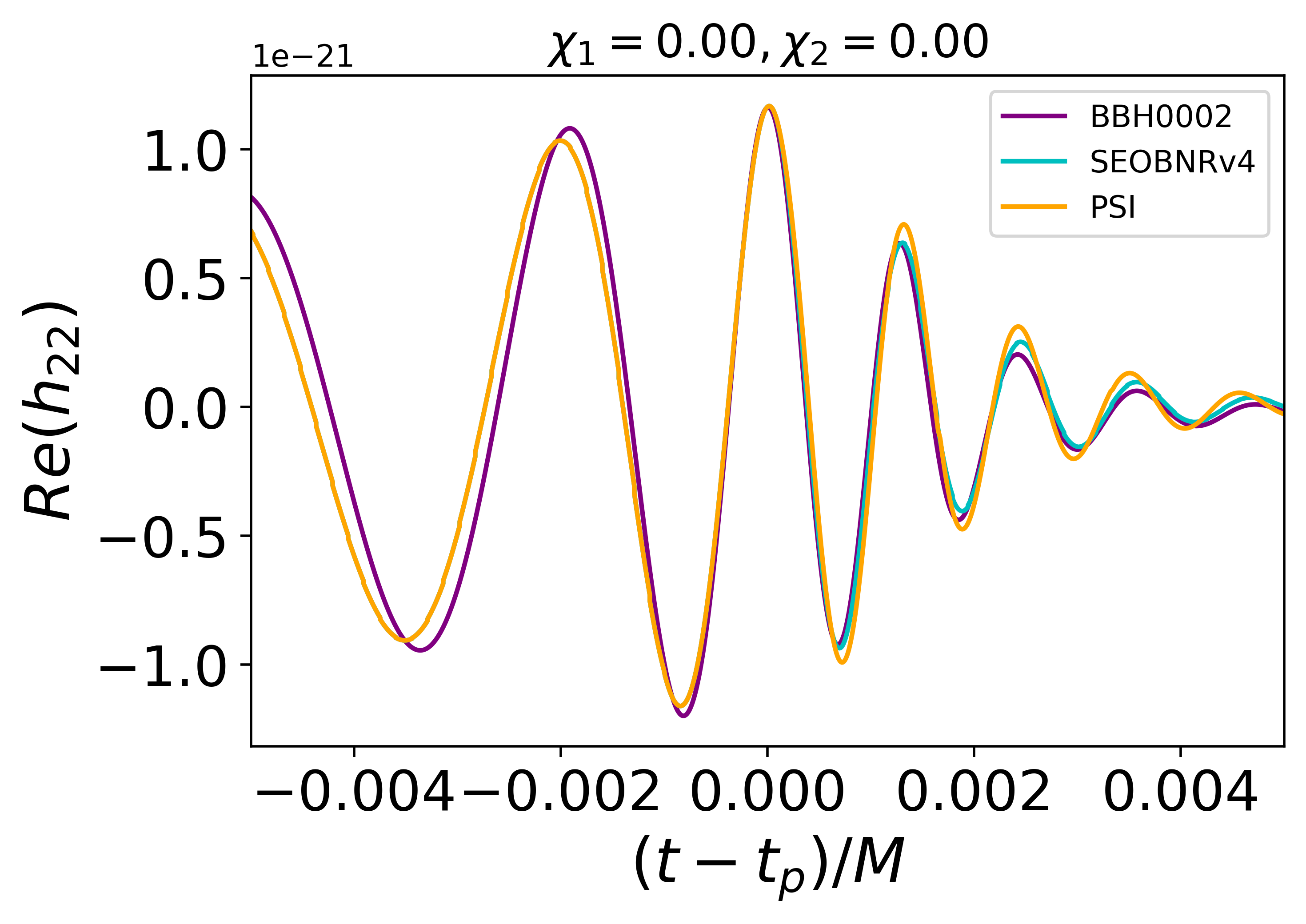}
\includegraphics[width=0.4 \textwidth]{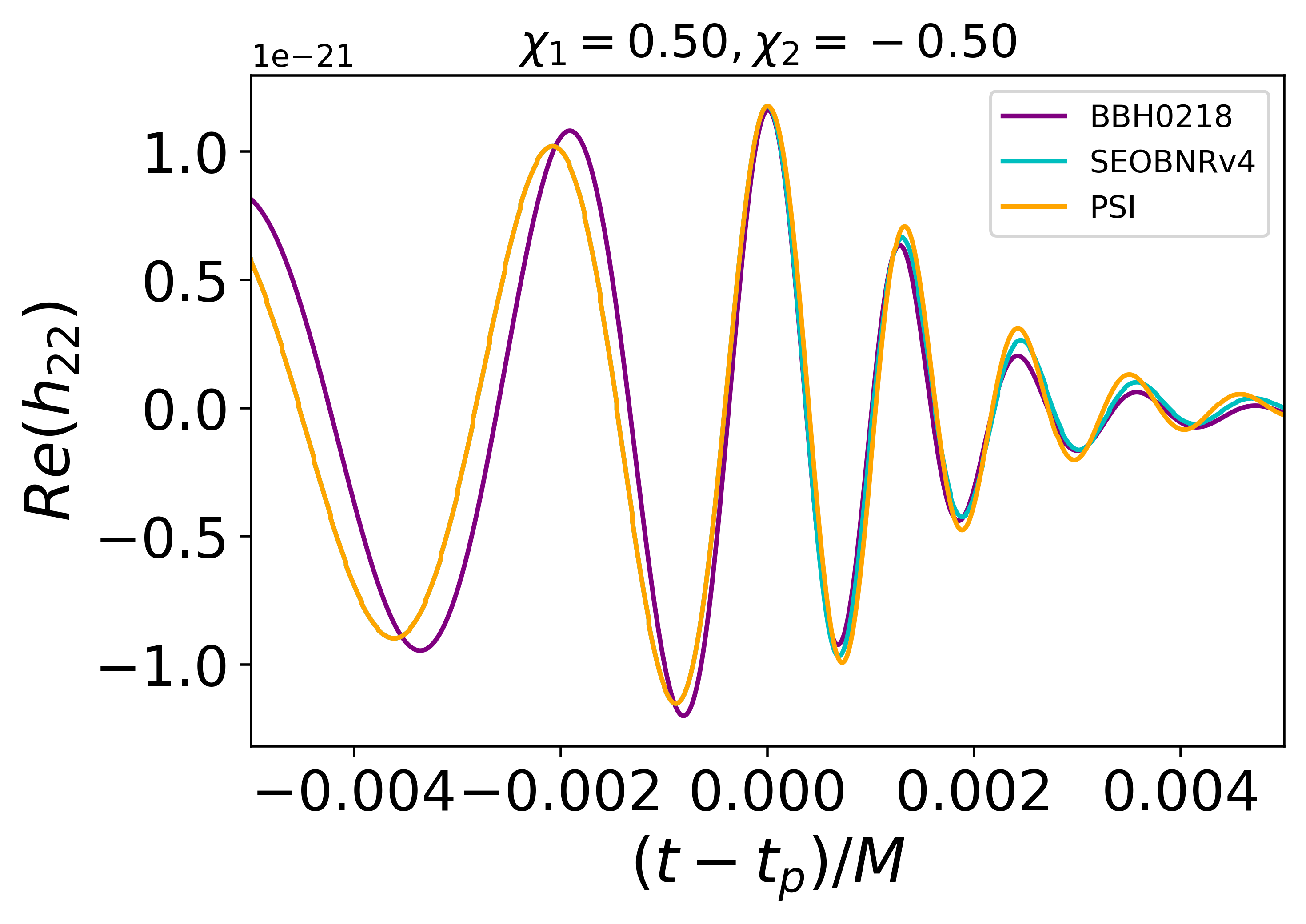}
\includegraphics[width=0.4 \textwidth]{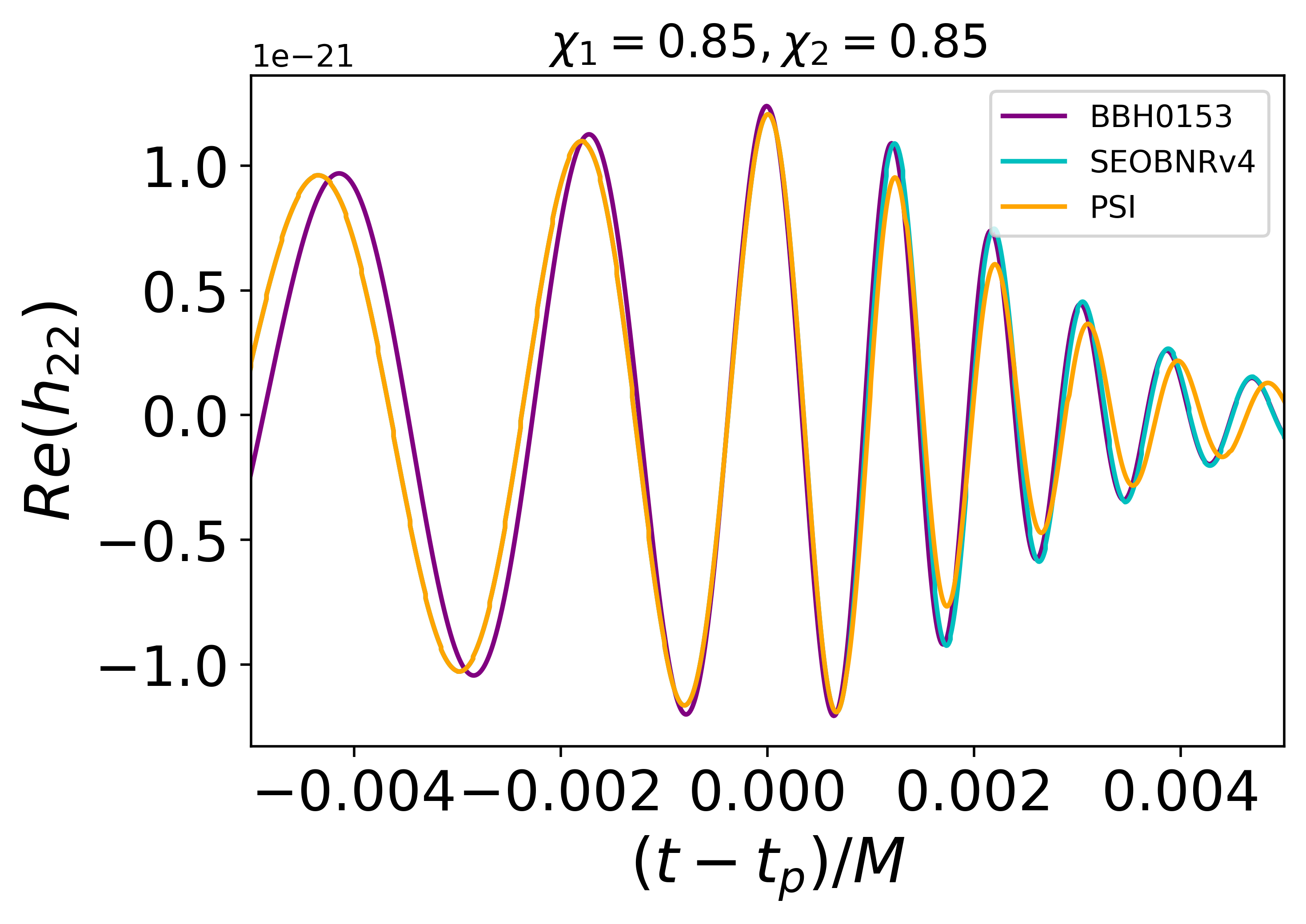}
\includegraphics[width=0.4 \textwidth]{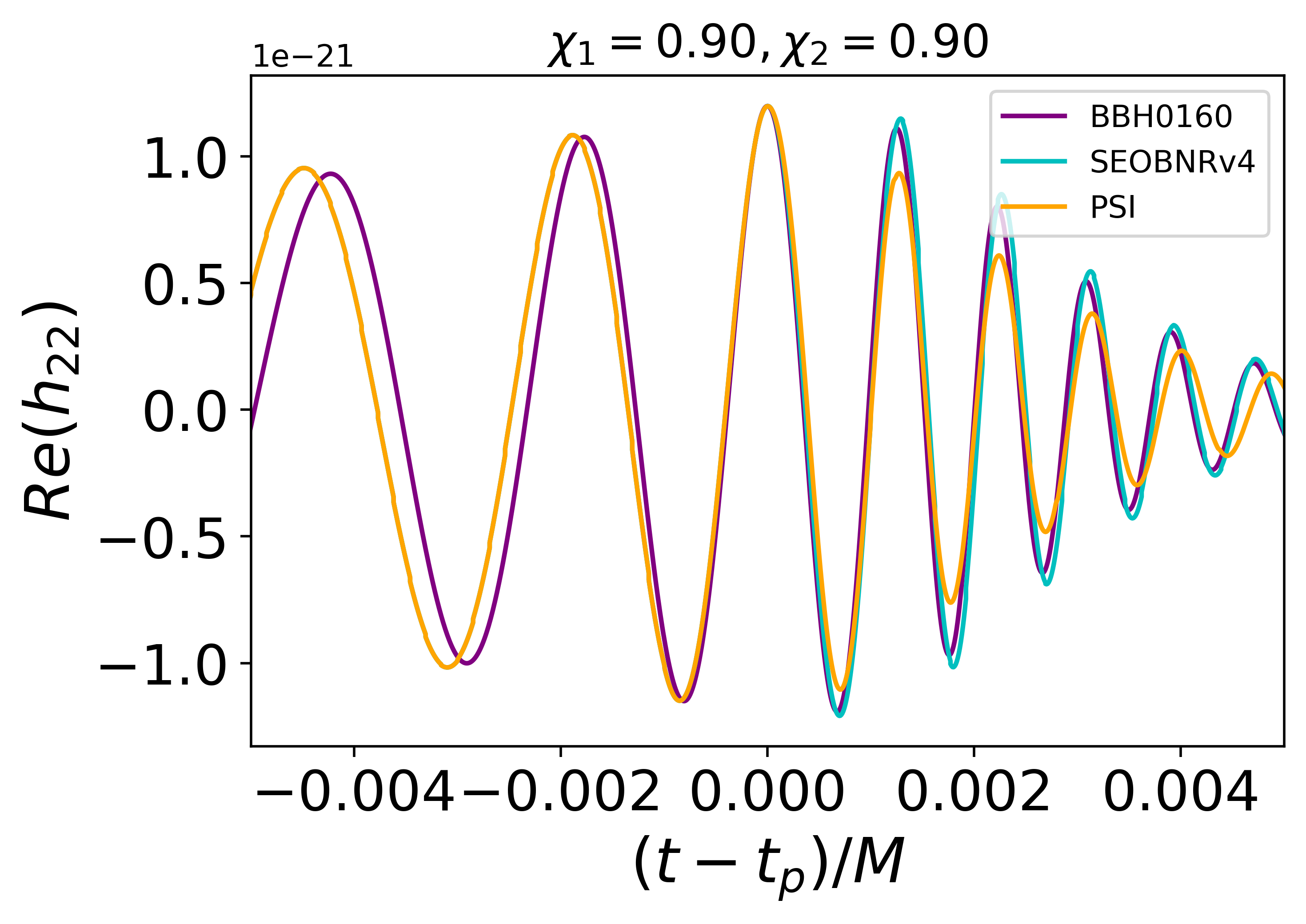}
\caption{Comparison of PSI waveforms with SEOBNRv4 and SXS for binary black holes systems with different spins. $\chi_{1}$ and $\chi_{2}$ represents the dimensionless spin of the first and second black hole, respectively. \label{compare2}}
\end{figure*}

Now we compare the PSI waveforms with NR data. We use the Simulating eXtreme Spacetimes (SXS) Collaboration catalog\cite{SXS_a, SXS_b} as the NR waveform data. With appropriate SXS data, we compare PSI, SEOBNRv4, and NR waveforms with different spins, as shown in Fig.~\ref{compare2}. We observe that the match between the PSI and NR waveforms is reduced for larger spin values. However, even for the dimensionless spins $\chi_{1,2}$ both equal to 0.9, the overlap is still larger than 98\%. Therefore the accuracy of PSI ($\Psi$) waveforms should be enough based on the comparison with NR data. This result implies that the $\Psi$ waveform model should be helpful in the GW data analysis.

\section{The ringdown waveforms from KRZ black holes}  \label{KRZ}
The KRZ metric was proposed by Konoplya, Rezzolla, and Zhidenko~\cite{KRZ}, who use the spacetime parametrization of generally axisymmetric black holes. The most significant difference between the KRZ metric and other metrics is that it contains many significant parameters, and each parameter is associated with the physical properties of the black hole. Therefore the KRZ metric can be used to describe the spacetime of various black holes in alternative gravity theories. The ringdown signals reflect the properties of black holes and can be used to test the nature of the black hole and then general relativity. We now use the photon sphere to construct the ringdown waveforms for the generally axisymmetric black holes. The lowest-order metric expression of the KRZ parametrization has the following form:

\begin{eqnarray}\label{metric}
d s^{2}&=&-\frac{N^{2}(\tilde{r}, \theta)-W^{2}(\tilde{r}, \theta) \sin ^{2} \theta}{K^{2}(\tilde{r}, \theta)} d t^{2}\nonumber\\&&-2 W(\tilde{r}, \theta) \tilde{r} \sin ^{2} \theta d t d \phi  \nonumber\\&&+K^{2}(\tilde{r}, \theta) \tilde{r}^{2} \sin ^{2} \theta d \phi^{2}\nonumber\\&&+\Sigma(\tilde{r}, \theta)\left(\frac{B^{2}(\tilde{r}, \theta)}{N^{2}(\tilde{r}, \theta)} d \tilde{r}^{2}+\tilde{r}^{2}d \theta^{2}\right), 
\end{eqnarray}

where $\tilde{r}=r/M, \ \tilde{a}=a / M$, and the other metric functions are defined as~\cite{xin2019}

\begin{eqnarray}
\Sigma&=&1+a^{2} \cos ^{2} \theta/\tilde{r}^{2}\ ,\\
{N}^{2}&=&\left( 1-{{r}_{0}}/\tilde{r} \right)\nonumber\\&&\left[ 1-{{\epsilon }_{0}}{{r}_{0}}/\tilde{r}+\left( {{k}_{00}}-{{\epsilon }_{0}} \right)r_{0}^{2}/{{{\tilde{r}}}^{2}}+{{\delta }_{1}}r_{0}^{3}/{{{\tilde{r}}}^{3}} \right]\nonumber\\&&+[ {{a}_{20}}r_{0}^{3}/{{{\tilde{r}}}^{3}}+{{a}_{21}}r_{0}^{4}/{{{\tilde{r}}}^{4}}+{{k}_{21}}r_{0}^{3}/{{{\tilde{r}}}^{3}}L]{{\cos }^{2}}\theta\ ,\\
B&=&1+\delta_{4} r_{0}^{2} / \tilde{r}^{2}+\delta_{5} r_{0}^{2} \cos ^{2} \theta / \tilde{r}^{2}\ ,\\
W&=&\left[w_{00} r_{0}^{2} / \tilde{r}^{2}+\delta_{2} r_{0}^{3} / \tilde{r}^{3}+\delta_{3} r_{0}^{3} / \tilde{r}^{3} \cos ^{2} \theta\right] / \Sigma\ ,\\
K^{2}&=&1+a W / r\nonumber\\&&+\left\{k_{00} r_{0}^{2} / \tilde{r}^{2}+k_{21} r_{0}^{3} / \tilde{r}^{3}L \cos ^{2} \theta\right\} / \Sigma\ ,
\\ \label{L}
L&=&\left[1+\frac{k_{22}\left(1-r_{0} / \tilde{r}\right)}{1+k_{23}\left(1-r_{0} / \tilde{r}\right)}\right]^{-1}\ .
\end{eqnarray}

In this paper we use the following parameters defined
as~\cite{xin2019}

\begin{eqnarray}
r_{0}&=&1+\sqrt{1-\tilde{a}^{2}}, \\  
a_{20}&=&2 \tilde{a}^{2} / r_{0}^{3}, \\ 
a_{21} &=&-\tilde{a}^{4} / r_{0}^{4}+\delta_{6}, \\ \epsilon_{0}&=&\left(2-r_{0}\right) / r_{0}, \\ k_{00}&=&\tilde{a}^{2} / r_{0}^{2} ,\\ 
k_{21}&=&\tilde{a}^{4} / r_{0}^{4}-2 \tilde{a}^{2} / r_{0}^{3}-\delta_{6}, \\ 
w_{00}&=& 2 \tilde{a} / r_{0}^{2}, \\ 
k_{22}&=&-\tilde{a}^{2} / r_{0}^{2}+\delta_{7}, \\ k_{23}&=&\tilde{a}^{2} / r_{0}^{2}+\delta_{8},
\end{eqnarray}
where $r_{0}$ is the radius of the event horizon in the equatorial plane and $\delta_{i}$\ ($i=1,2,3,4,5,6,7,8$) is the dimensionless parameter describing the corresponding deformation of the parameter in the metric~(\ref{metric}). Particularly, $\delta_{1}$ corresponds to the deformation of $g_{tt}$, $\delta_{2}$, and $\delta_{3}$ refer to the deformations of spin, $\delta_{4}$ and $\delta_{5}$ relate to the deformations of $g_{rr}$, $\delta_{6}$ is for the deformation of the event horizon. In the case where $\delta_{i}=0$, the KRZ one~(\ref{metric}) reduces to the Kerr metric, and $\tilde{a}=0$ reduces the Kerr metric to the Schwarzschild one.

\begin{figure*}
\centering
\subfigure[Photon circular orbital radius($R$)]{
\includegraphics[width=0.4 \textwidth]{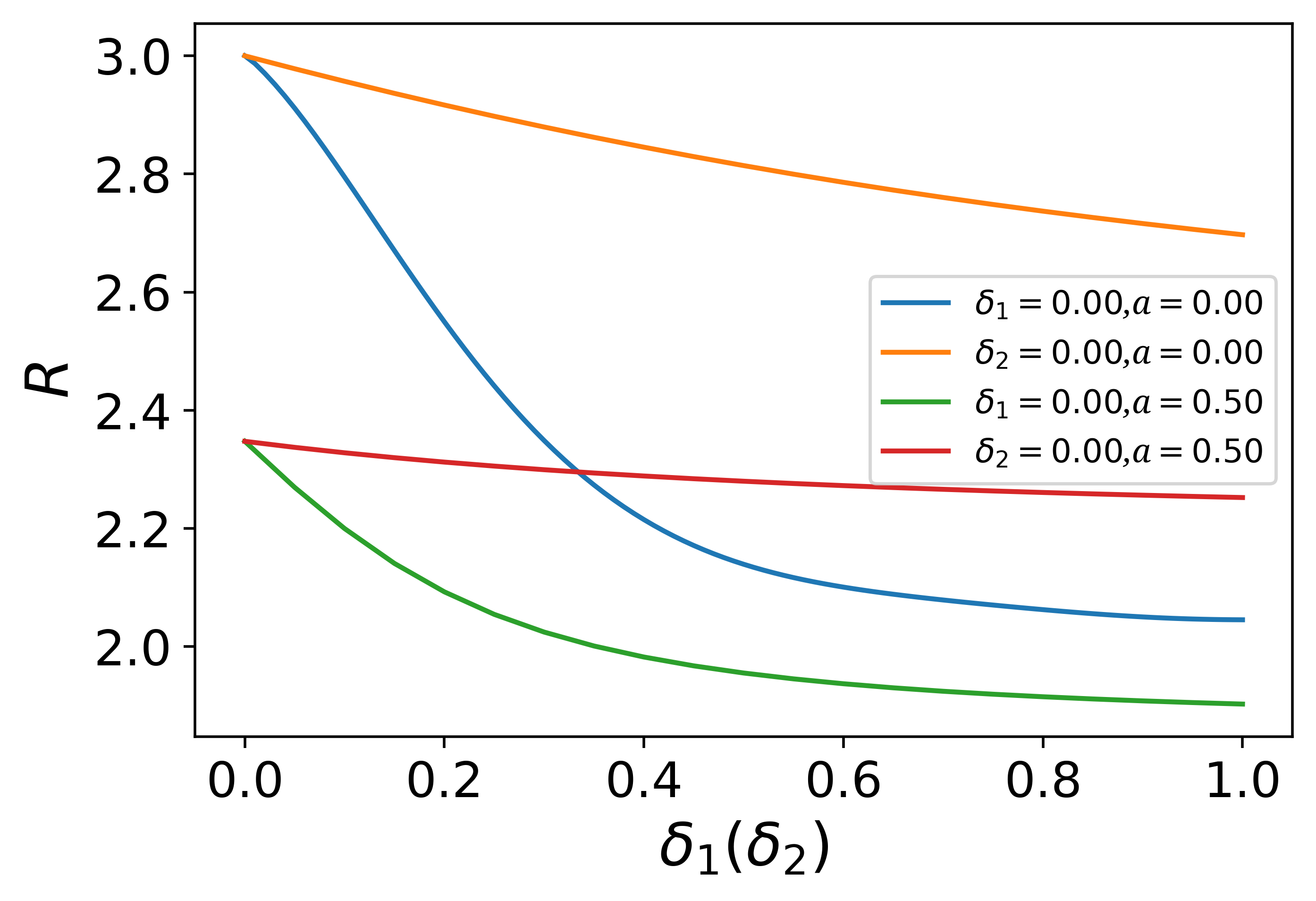}
}
\quad
\subfigure[Frequency of GWs($\omega_R$)]{
\includegraphics[width=0.4 \textwidth]{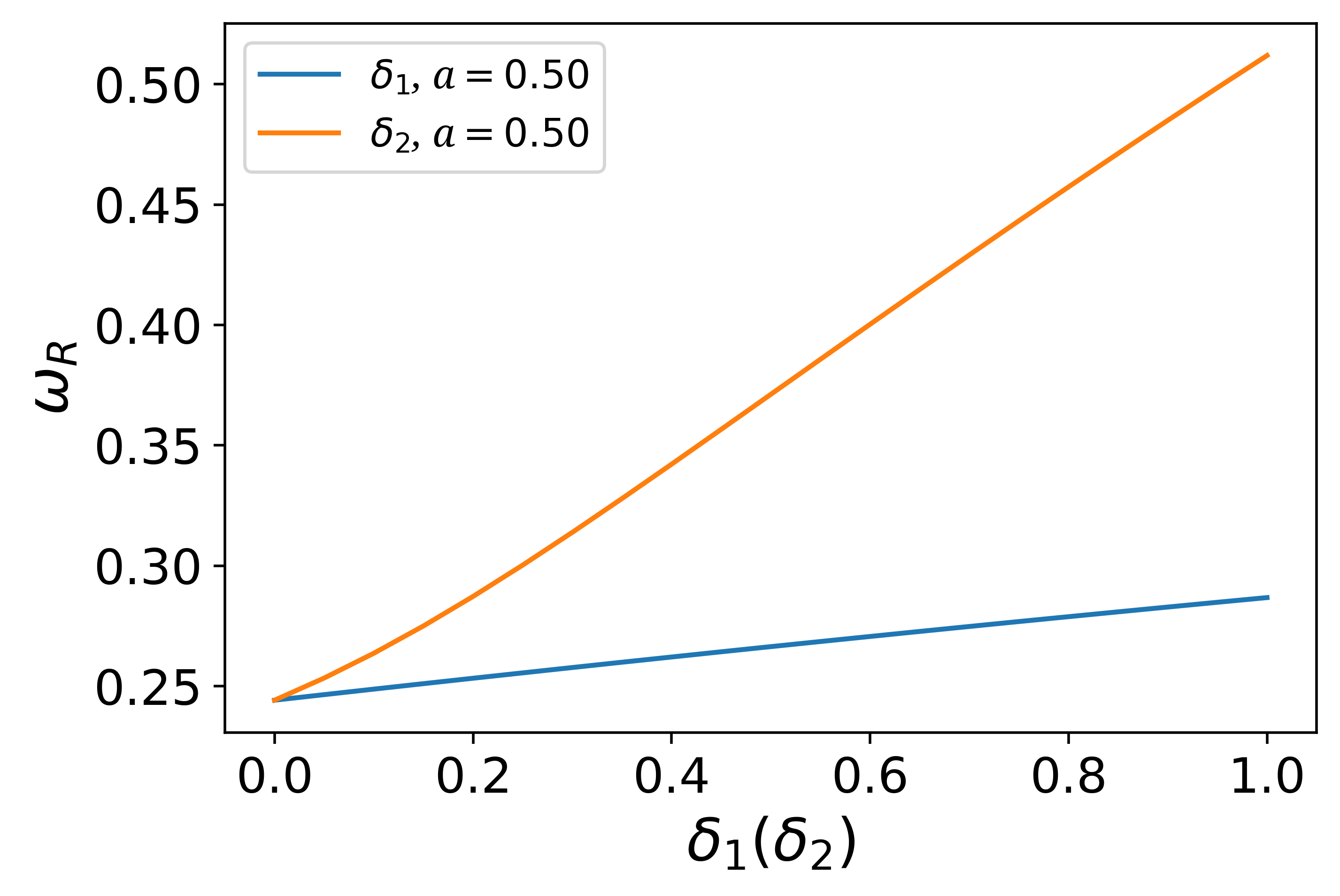}
}
\caption{The photon circular orbital radius and $\omega_R$ with different values of $\delta_1(\delta_2)$ in the equatorial plane under the KRZ metric. The left figure shows the effect of $\delta_1$ or $\delta_2$ on the photon circular orbital radius with two different spin cases: $a=0.00$ and $a=0.50$. The right figure shows the effect of $\delta_1$ or $\delta_2$ on the $\omega_R$ under spin $a=0.50$. It should be noted that when we study the influence of one of the $\delta$, the other $\delta$ we take as $0$. \label{KRZ_R2}}
\end{figure*}

To get the ringdown waveforms of perturbed KRZ black holes, we need the radius of the circular photon orbits and the values of $\omega_R$ and $\omega_I$. The way to calculate the radius of the circular photon orbits, $\omega_R$ and $\omega_I$, is given in this paper\cite{Li2021}. This metric only retains three dimensionless parameters $\delta_1$, $\delta_2$, and $\delta_4$ when we consider the equatorial plane and for simplicity we set $\delta_4=0.00$ in this paper. The photon orbits and $\omega_R$ are shown in Fig.~\ref{KRZ_R2}. We observe in Fig.~\ref{KRZ_R2}(a) that the radius of the photon sphere shrinks as $\delta_1(\delta_2)$ increases, and the effect of $\delta_2$ is more obvious than $\delta_1$. With the rise of spin $a$, the decrease in radius is weakened. We observe in Fig.~\ref{KRZ_R2}(b) that $\delta_2$ has more influence on $\omega_R$, which also means that $\delta_2$ has a more significant impact on ringdown waveforms.

\begin{figure*}
\centering
\subfigure[$a$ = 0.50 and $\delta_2=0$]{
\includegraphics[width=0.4 \textwidth]{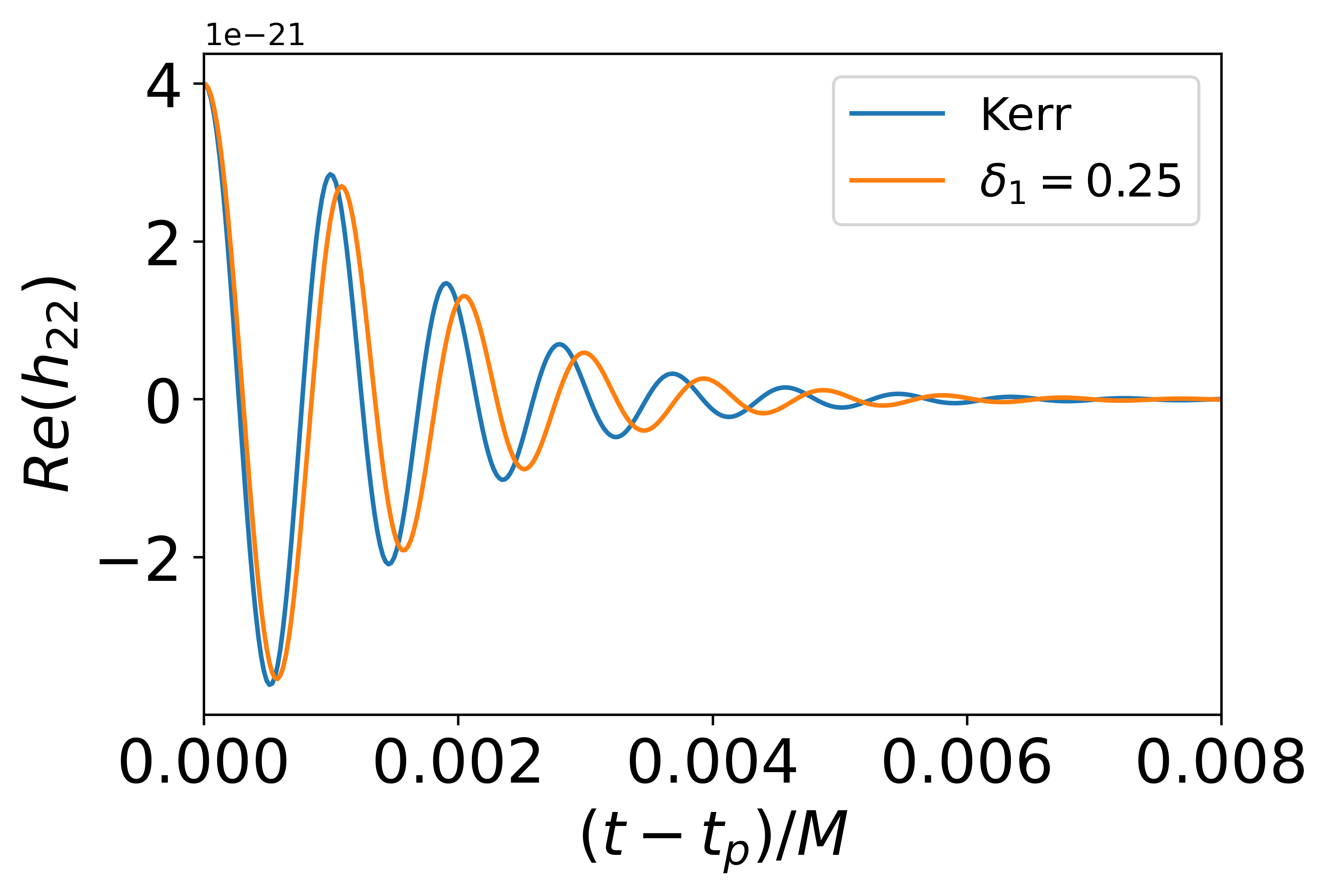}
}
\quad
\subfigure[$a$ = 0.50 and $\delta_2=0$]{
\includegraphics[width=0.4 \textwidth]{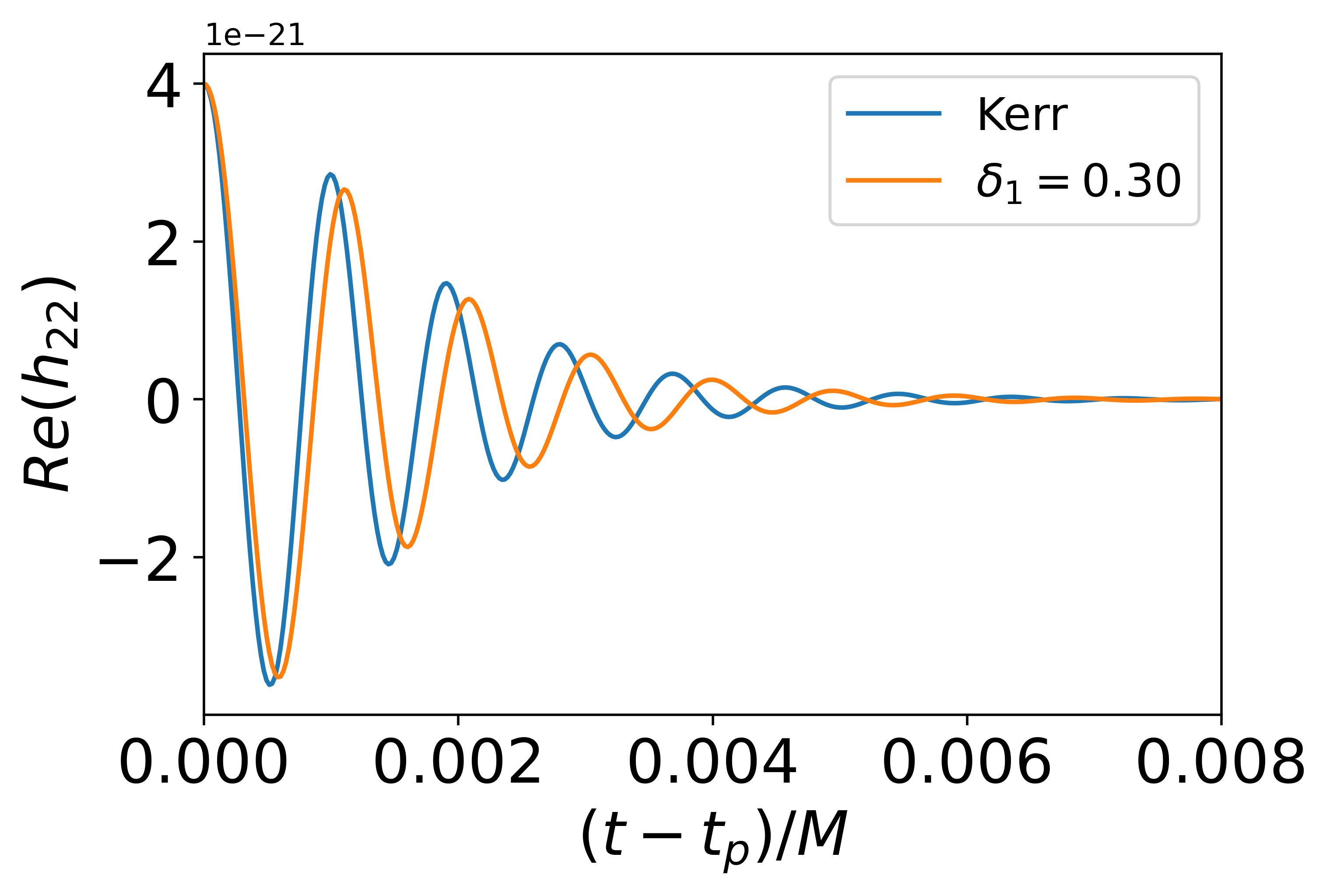}
}
\quad
\subfigure[$a$ = 0.50 and $\delta_1=0$]{
\includegraphics[width=0.4 \textwidth]{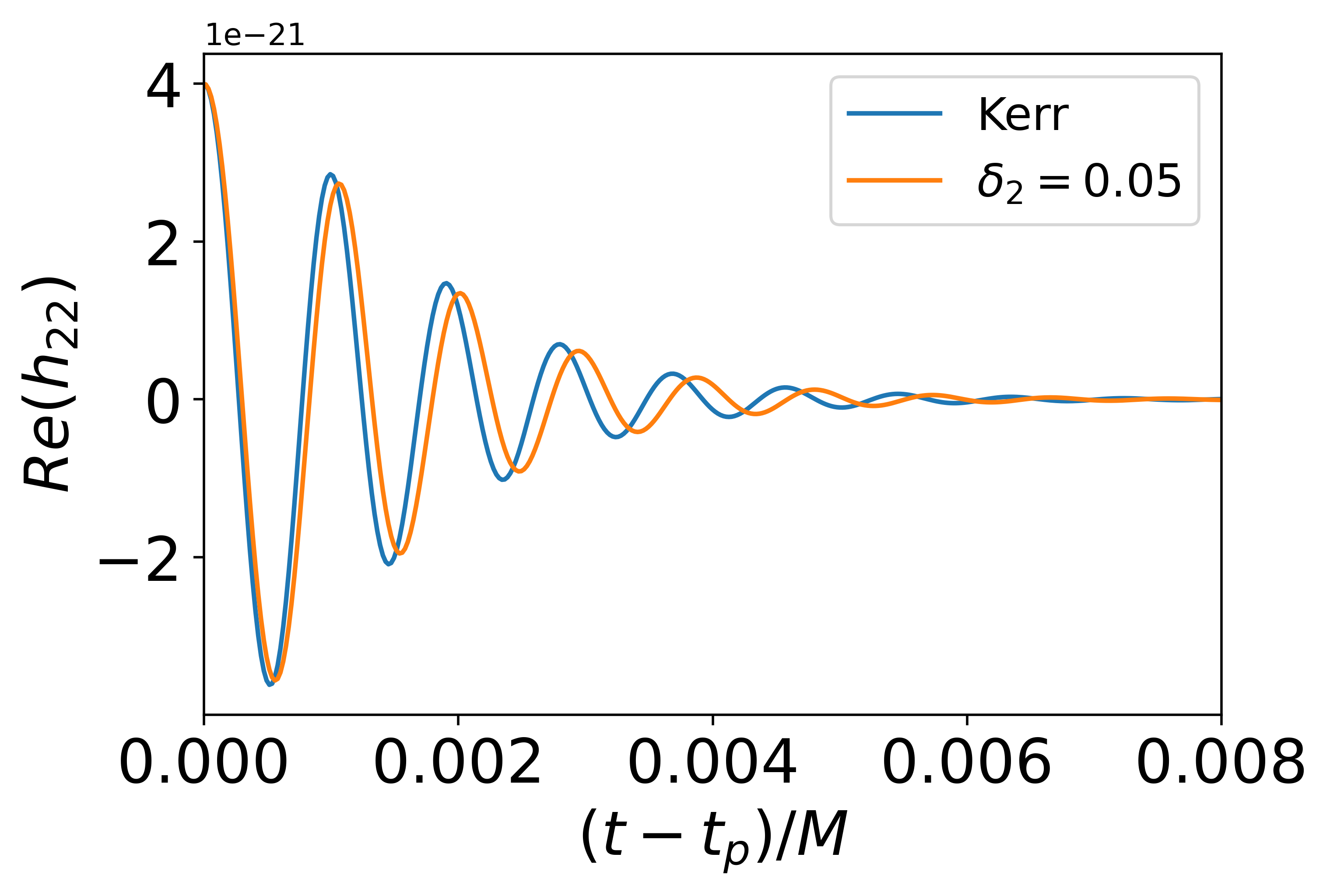}
}
\quad
\subfigure[$a$ = 0.50 and $\delta_1=0$]{
\includegraphics[width=0.4 \textwidth]{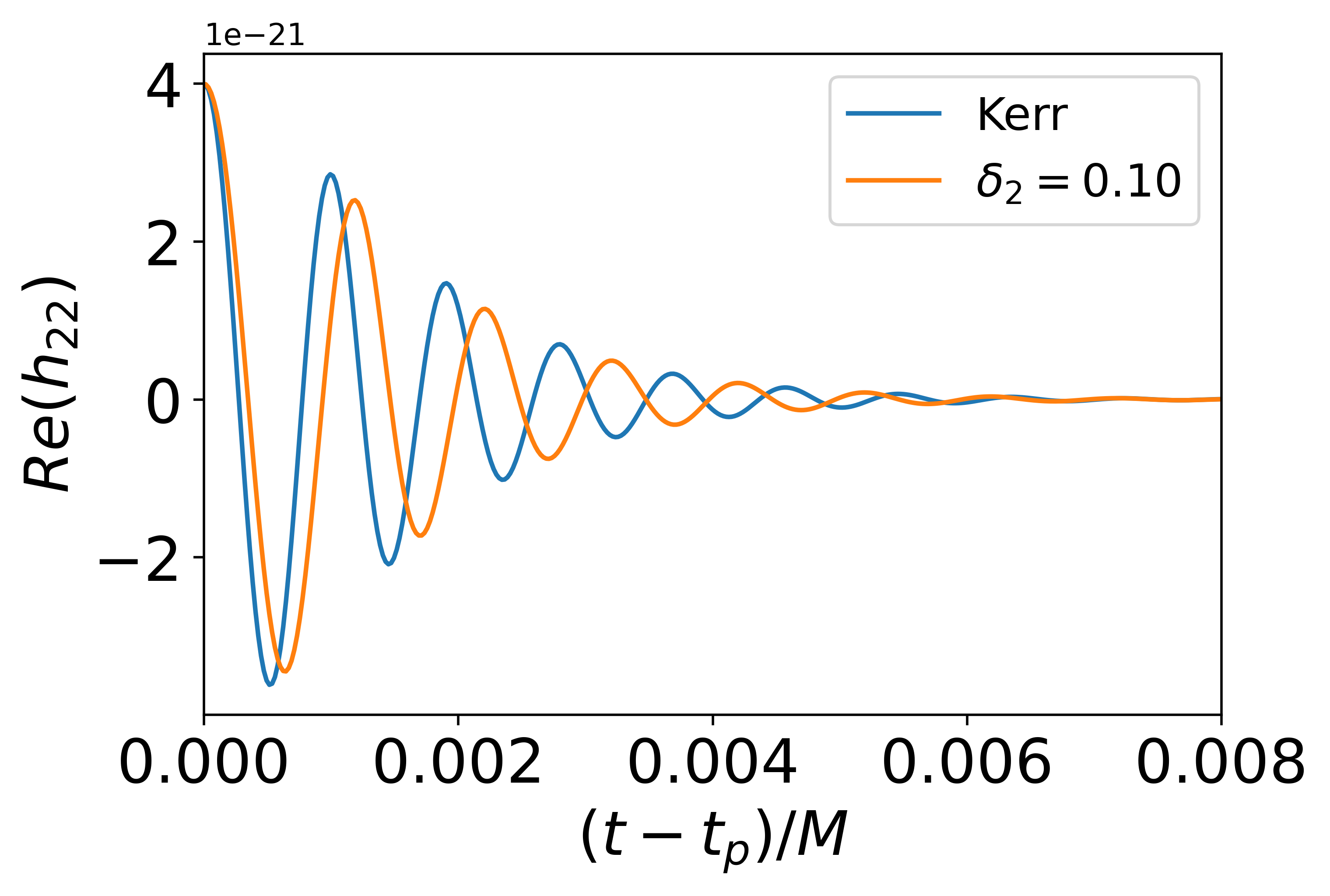}
}
\caption{The ringdown waveform of perturbed KRZ black holes with various values of $\delta_1$ and $\delta_2$. The top panels show the ringdown waveform with various $\delta_1$(fixing $\delta_2=0.00$ and $a=0.50$); The bottom panels show the ringdown waves with various $\delta_2$(fixing $\delta_1=0.00$ and $a=0.50$).\label{KRZ_GW}}
\end{figure*}

Then we can get the ringdown signals as shown in Fig.~\ref{KRZ_GW}. From the first and second rows, with the increase of $\delta_1$($\delta_2$), the waveforms can be easily distinguished from the Kerr cases, and the influence of $\delta_2$ is more significant than $\delta_1$. The result is reasonable because $\delta_2$ has a more substantial effect in $\omega_R$ than $\delta_1$ as shown in Fig.~\ref{KRZ_R2}.

\begin{figure*}
\centering
\subfigure[$a$ = 0.50 and $\delta_2=0$]{
\includegraphics[width=0.4 \textwidth]{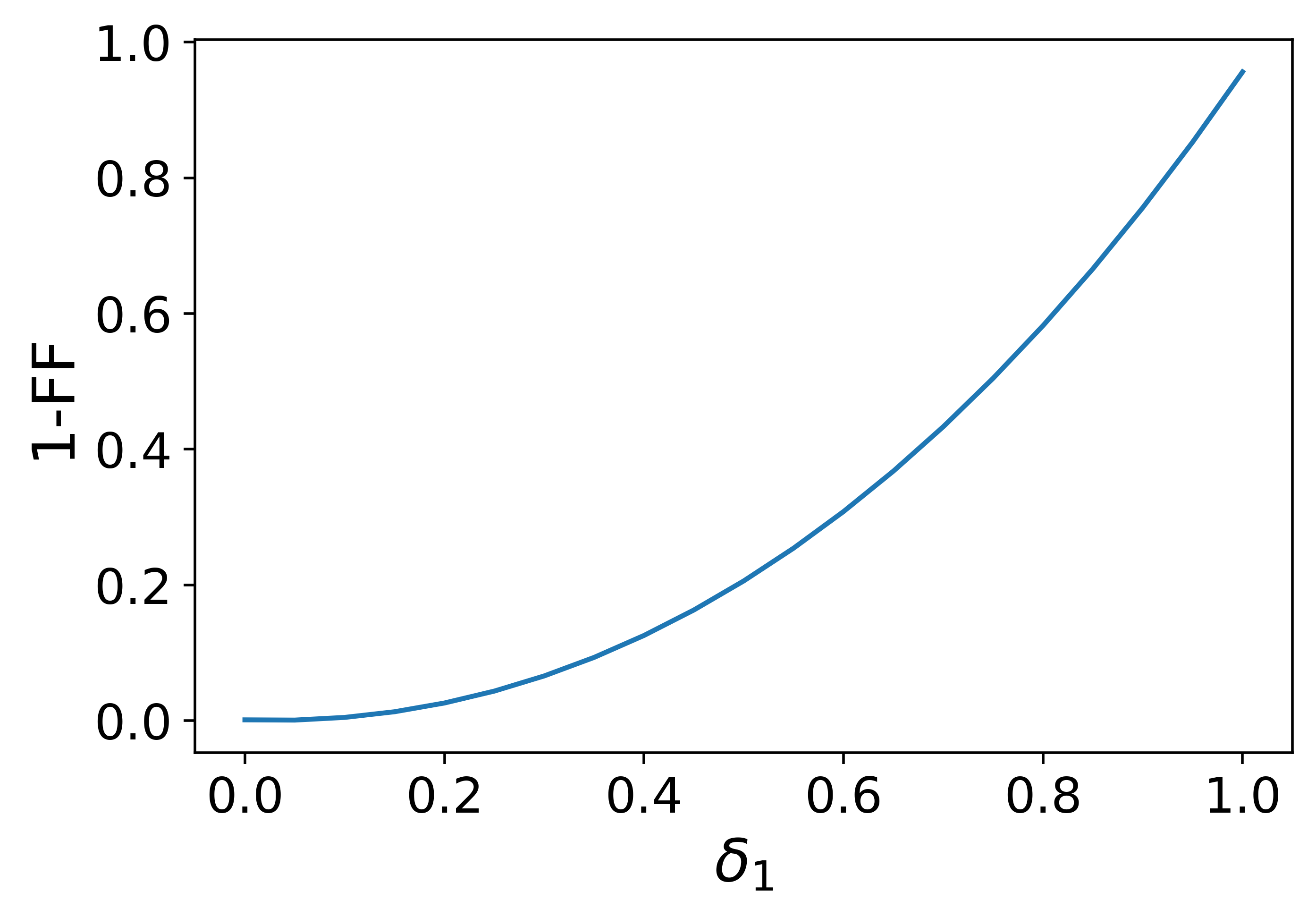}
}
\quad
\subfigure[$a$ = 0.50 and $\delta_1=0$]{
\includegraphics[width=0.4 \textwidth]{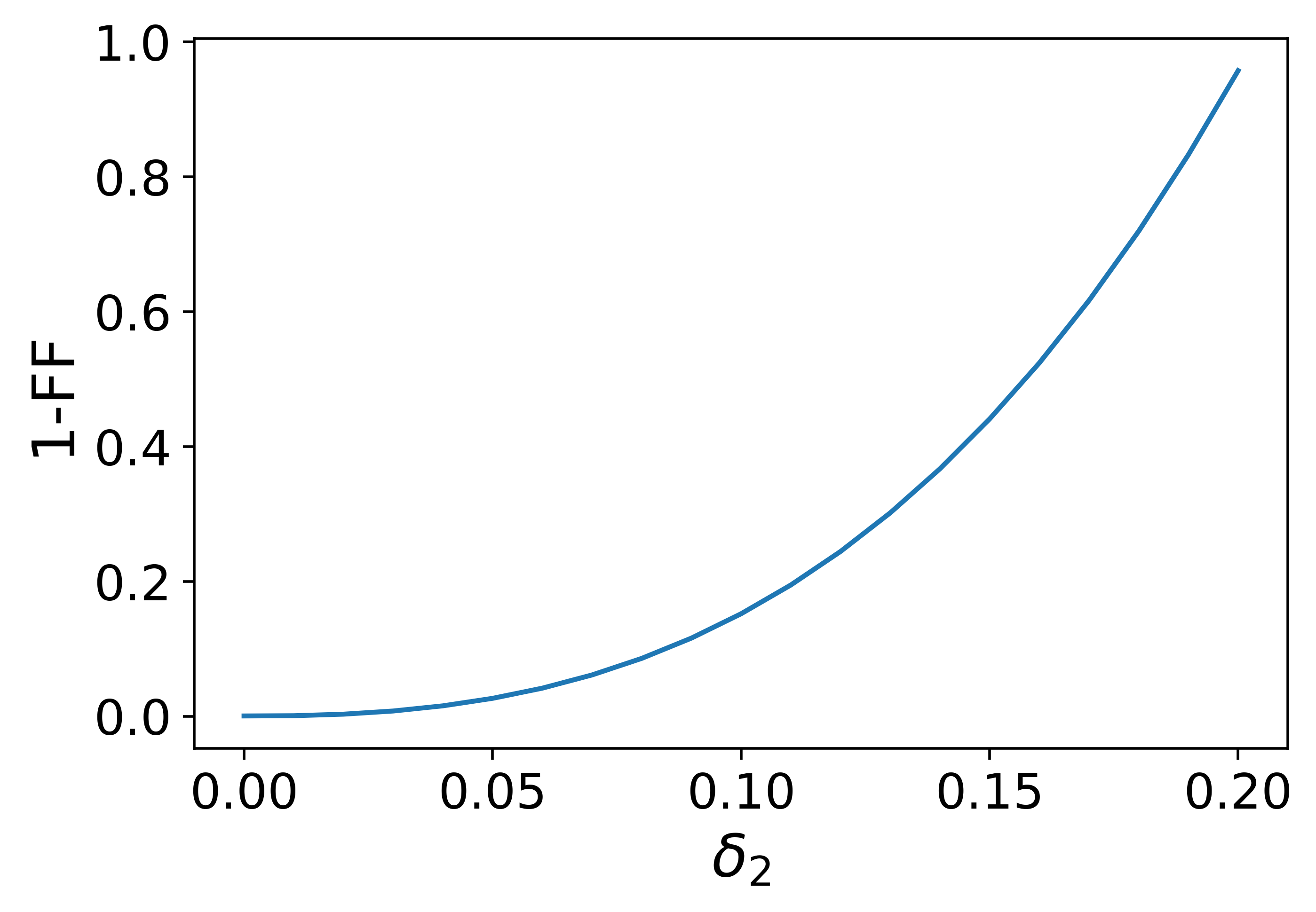}
}
\caption{The mismatch (1-FF) of ringdown waves between the Kerr and KRZ black holes with varied parameters $\delta_1$ and $\delta_2$.
\label{KRZ_Mismatch}}
\end{figure*}

We also study the mismatch (1-overlap) of gravitational-wave waveforms between the Kerr black hole and KRZ black hole to see the influence of the two parameters $\delta_1$ and $\delta_2$. Through Fig.~\ref{KRZ_Mismatch}, it is clear that with the increase of $\delta_1$($\delta_2$) the mismatch increases. We observe that the influence of $\delta_2$ is more significant than $\delta_1$, when $\delta_1=1.00$ the mismatch equals to $1.0$, but when $\delta_2=0.20$ the mismatch already equals to $1.0$. The other KRZ parameters could be investigated in future work.

\section{Ringdown from accelerating black holes}\label{ABH}
The forming black hole can be accelerated to a large velocity due to anisotropically radiating GWs during the binary black hole merger. This phenomenon is gravitational recoil, and the kick velocity can be as large as a few thousand kilometers per second in a very short duration ($\sim 20 M$).

The kick velocity will induce a redshift or blueshift on the GWs, which is already investigated in literature (see \cite{Favata_04, Blanchet_14, ABH_a, Torres_20, Han_19} and references inside). In addition, the acceleration should produce some additional effects, which are rarely discussed. In this section, we want to use the PSI model to calculate the ringdown signals from accelerating black holes and evaluate whether the GWs can detect this recoil acceleration directly. The metric of the accelerating Kerr black hole is\cite{ABH_b}

\begin{equation}
\begin{aligned}
d s^{2}=& \frac{1}{\Omega^{2}}\left[\Sigma\left(\frac{d \theta^{2}}{\Delta_{\theta}}+\frac{d r^{2}}{\Delta_{r}}\right)-\frac{\left(\Delta_{r}-a^{2} \Delta_{\theta} \sin ^{2} \theta\right) d t^{2}}{\Sigma}\right.\\
&+\frac{2\left[\chi \Delta_{r}-a \Delta_{\theta} \sin ^{2} \theta(a \chi+\Sigma)\right] d t d \phi}{\Sigma} \\
&\left.+\frac{\left.\left[\Delta_{\theta} \sin ^{2} \theta(a \chi+\Sigma)^{2}-\chi^{2} \Delta_{r}\right] d \phi^{2}\right]}{\Sigma}\right],
\end{aligned}
\end{equation}

where
\begin{equation}
\begin{aligned}
\chi &=a \sin ^{2} \theta\,, \\
\Omega &=1-A r \cos \theta\,, \\
\Sigma &=r^{2}+a^{2} \cos ^{2} \theta\,, \\
\Delta_{r} &=\left(1-A^{2} r^{2}\right)\left(r^{2}-2 m r+a^{2}\right)\,, \\
\Delta_{\theta} &=1-2 A m \cos \theta+a^{2} A^{2} \cos ^{2} \theta\,,
\end{aligned}
\end{equation}
where $a$, $m$ are the spin and mass of the BH. $A$ is the acceleration of the BH with a unit of $1/M$.

\begin{figure}
\includegraphics[width=0.5 \textwidth]{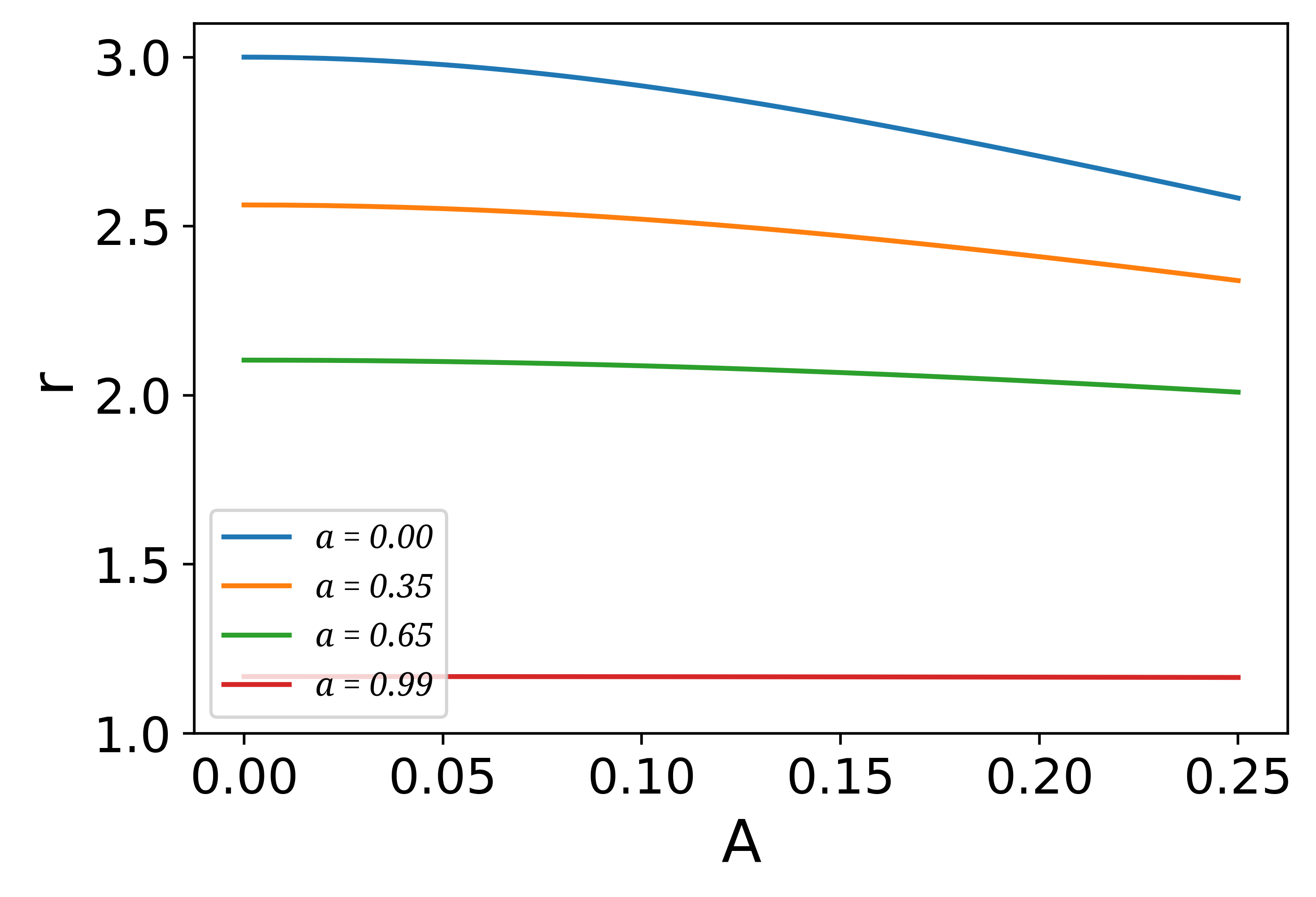}
\caption{The radius of the photon circular orbits for varied acceleration parameter $A$. \label{Radius_ABH}}
\end{figure}

First we calculate the radius of the photon circular orbits, for convenience, we use this equation\cite{H_a}:
\begin{equation}
H_{\pm}(r, \theta) \equiv \frac{-g_{t \phi} \pm \sqrt{D}}{g_{\phi \phi}}\,,
\end{equation}
where $D \equiv\left(g_{t \phi}^{2}-g_{t t} g_{\phi \dot{\phi}}\right)$, the photon circular orbits occur at $\frac{\partial {{H}_{\pm }}}{\partial r}=0$ and we only consider the prograde orbits(i.e., $H_{+}$). Through this method, we can get the radius of the circular photon orbits in Fig.~\ref{Radius_ABH}. It is shown that the change of radius of the photon sphere is more obvious for a small spin, so we use $a=0.01$ in the following study.

\begin{figure}
\includegraphics[width=0.5 \textwidth]{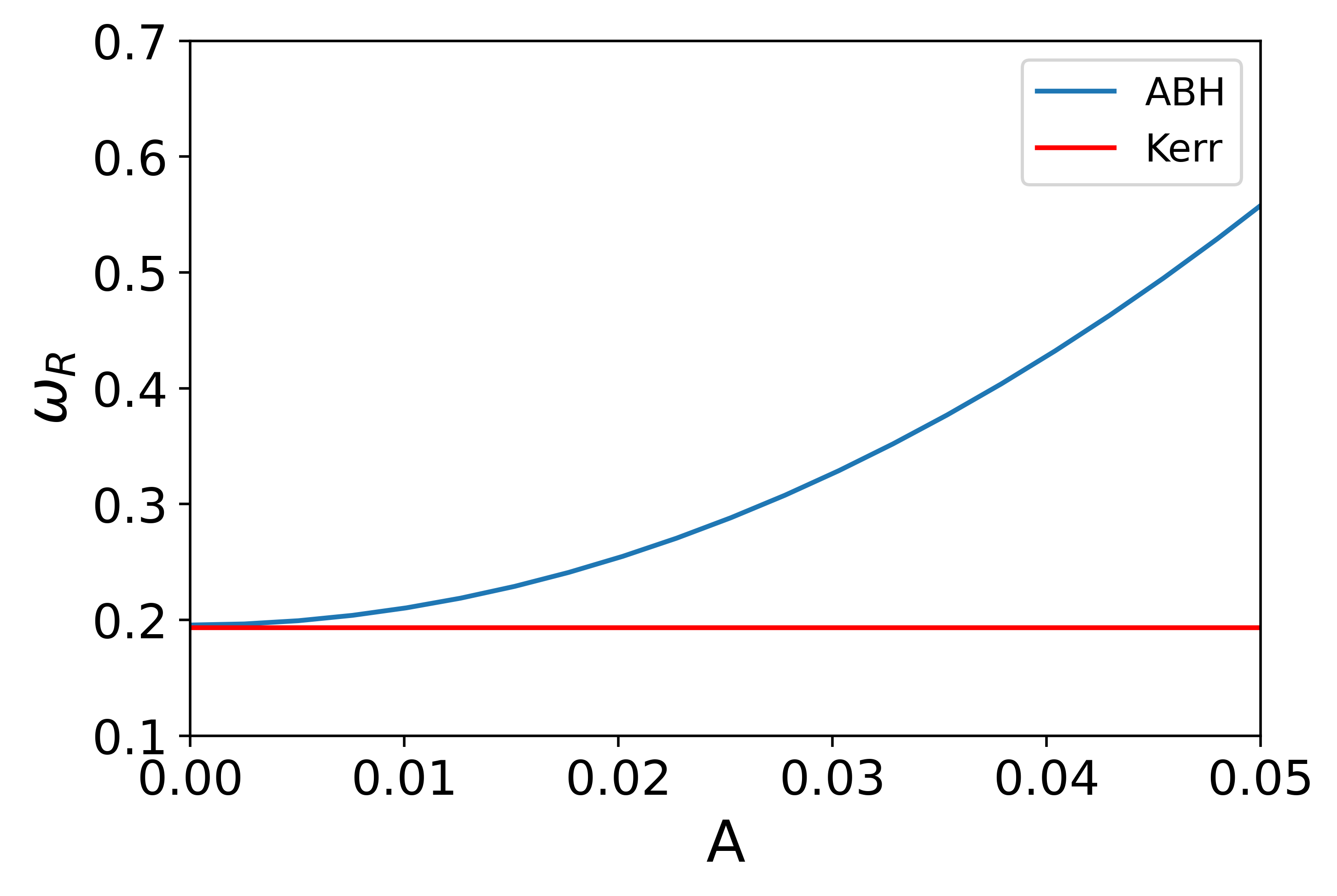}
\caption{The value of $\omega_R$ under accelerating black hole ($a=0.01$) versus acceleration $A$. \label{omega_R_ABH}}
\end{figure}

\begin{figure}
\includegraphics[width=0.5 \textwidth]{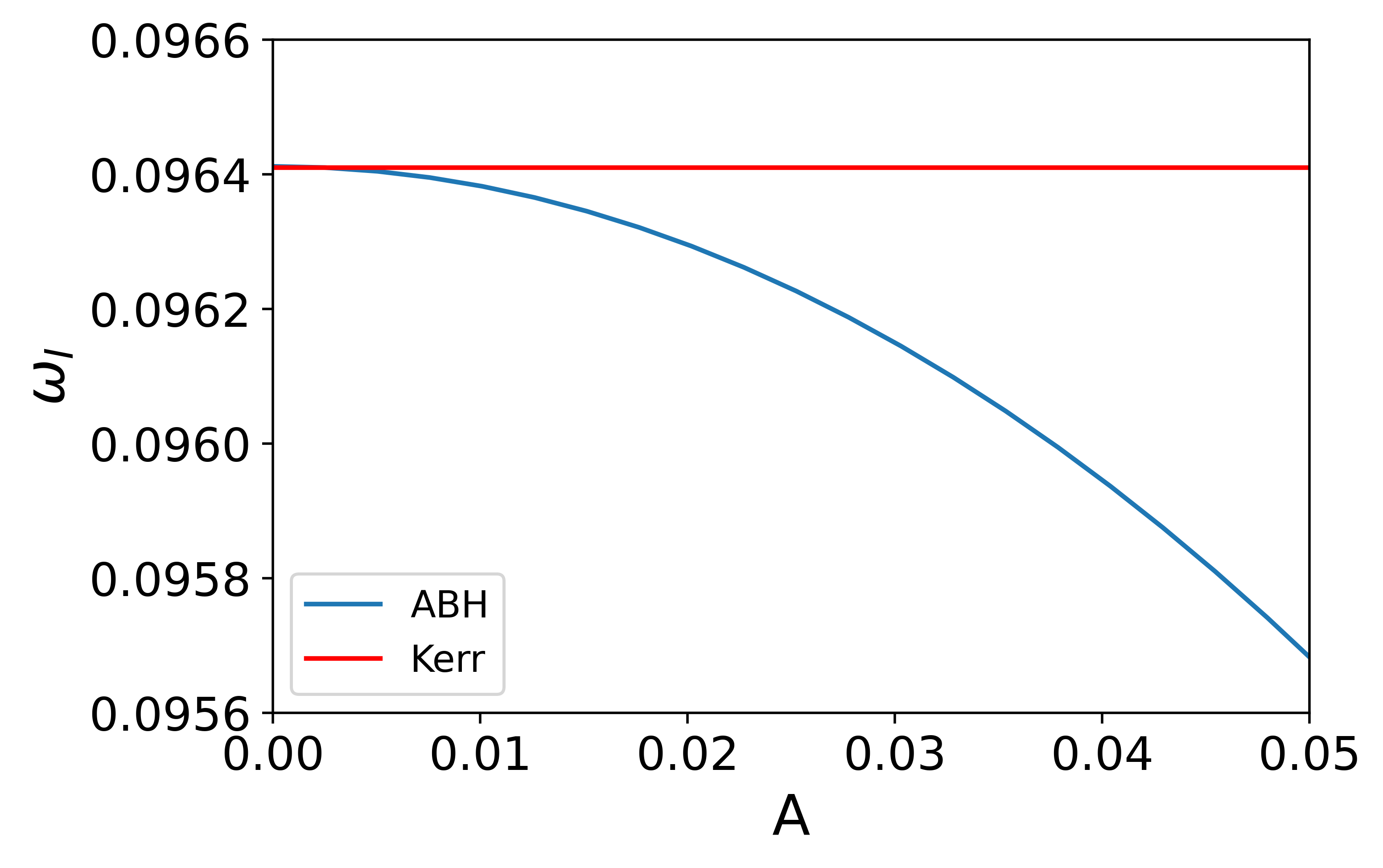}
\caption{The value of $\omega_I$ under accelerating black hole ($a=0.01$) versus acceleration $A$. \label{omega_I_ABH}}
\end{figure}

Before we calculate $\omega_R$ and $\omega_I$ for an accelerating BH, we should write the equations of the photon motion:
\begin{equation}
\begin{aligned}
&\frac{\Sigma}{\Omega^{2}} \frac{d t}{d \tau}=\frac{\chi\left(L_{z}-E \chi\right)}{\Delta_{\theta} \sin ^{2} \theta}+\frac{\left.(\Sigma+a \chi)[\Sigma+a \chi) E-a L_{z}\right]}{\Delta_{r}}\\
&\frac{\Sigma}{\Omega^{2}} \frac{d \phi}{d \tau}=\frac{L_{z}-E \chi}{\Delta_{\theta} \sin ^{2} \theta}+\frac{\left.a[\Sigma+a \chi) E-a L_{z}\right]}{\Delta_{r}},\\
&\left(\frac{\Sigma}{\Omega^{2}}\right)^{2}\left(\frac{d \theta}{d \tau}\right)^{2}=\Delta_{\theta} K-\frac{\left(\chi E-L_{z}\right)^{2}}{\sin ^{2} \theta} \equiv \Theta(\theta),\\
&\left(\frac{\Sigma}{\Omega^{2}}\right)^{2}\left(\frac{d r}{d \tau}\right)^{2}=\left[(\Sigma+a \chi) E-a L_{z}\right]^{2}-\Delta_{r} K \equiv R(r)
\end{aligned}
\end{equation}
where $\tau$ is the affine parameter, $E$ and $L_z$ are the photon's energy and angular momentum. Because we consider the circular photon orbits, these orbits will satisfy the conditions:$R\left(r_{p}\right)=0, \quad R^{\prime}\left(r_{p}\right)=0$, then we can get the value of $K$ and $L$. With the definition $\bar{K}=K/E^2$ and $\bar{L}=L/E$, we have
\begin{equation}
\bar{K}=\frac{1}{a}\left(\Sigma+a \chi-\frac{4 r \Delta_{r}}{\Delta_{r}^{\prime}}\right)\,,
\end{equation}
\begin{equation}
\bar{L}=\frac{1}{a}\left(\Sigma+a \chi-\frac{4 r \Delta_{r}}{\Delta_{r}^{\prime}}\right)\,.
\end{equation}
From these equations, we can get $\omega_R$ and $\omega_I$ as shown in Figs.~\ref{omega_R_ABH} and \ref{omega_I_ABH}. We observe in these figures that with the acceleration parameter $A$ increase, the difference of $\omega_R, \omega_I$ between the accelerating and Kerr BHs is more significant. Additionally, the influence of acceleration is more evident on the ringdown frequency ($\omega_R$) than on the decay rate ($\omega_I$).

\begin{figure*}
\centering
\includegraphics[width=0.4 \textwidth]{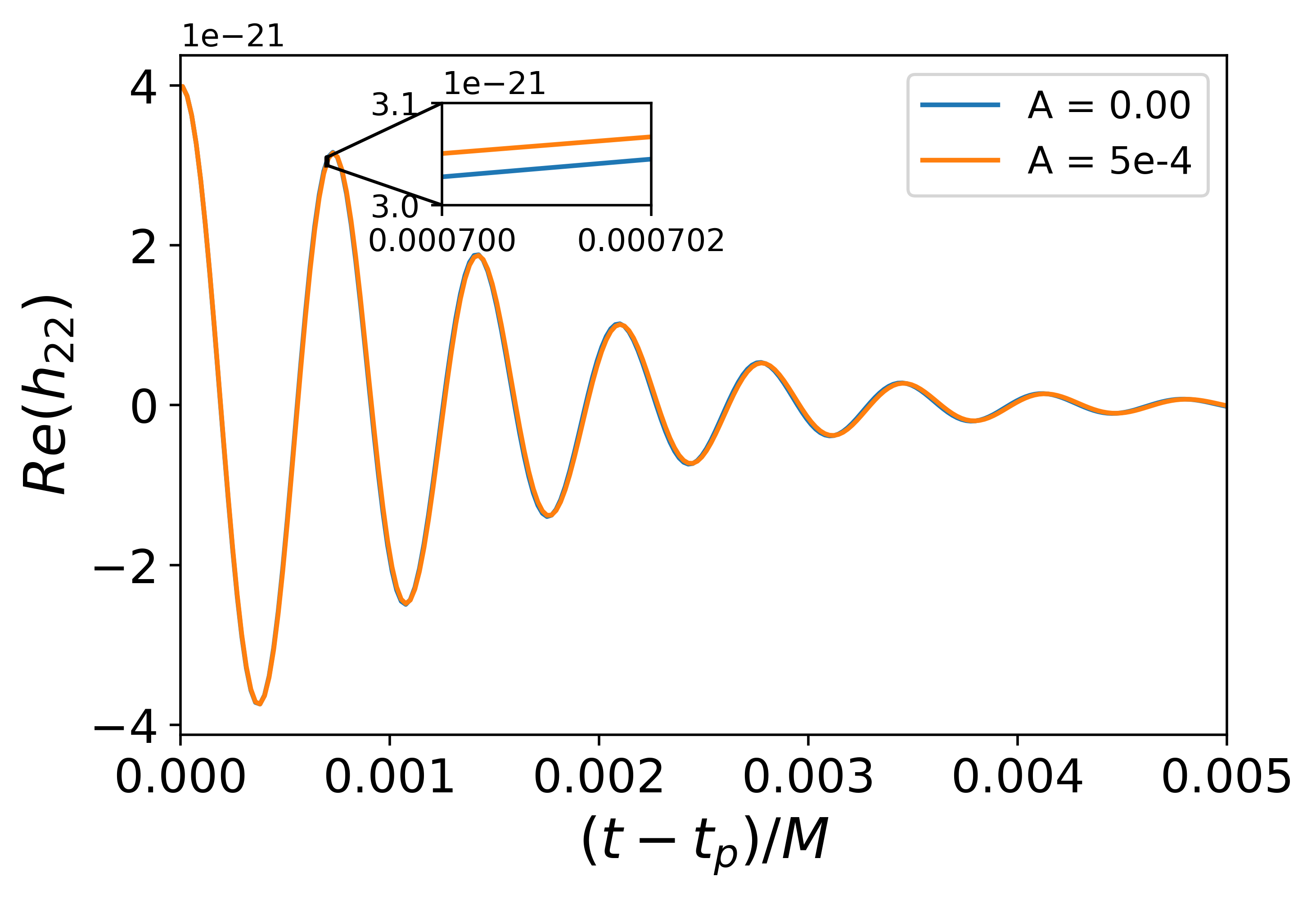}
\quad
\includegraphics[width=0.4 \textwidth]{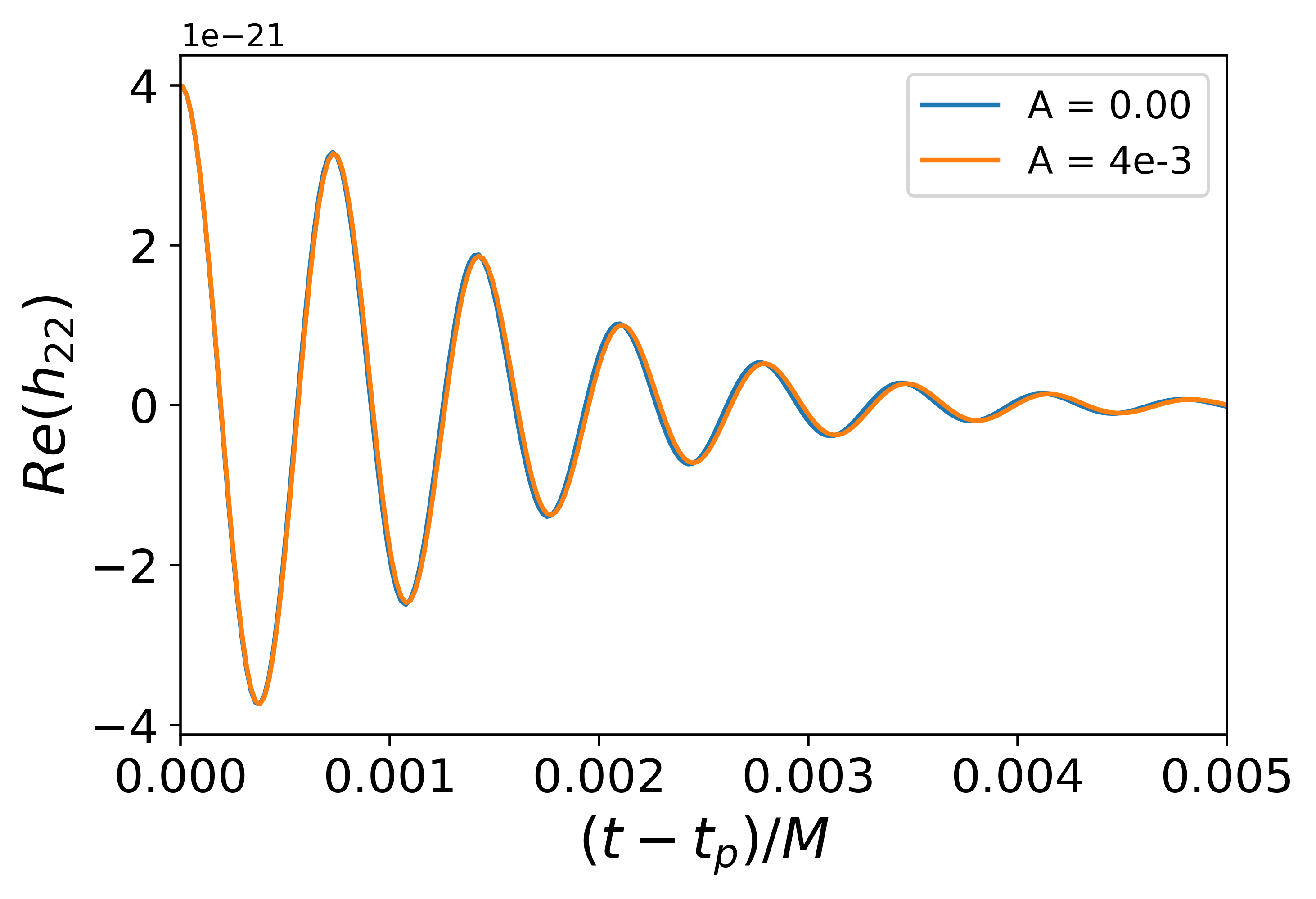}
\caption{The ringdown signals of accelerating and no-acceleration Kerr BHs ($a=0.01$). The left panel is the ringdown with $A = 5\times10^{-4}$, and the right panel shows the results for $A = 4\times10^{-3}$. \label{GW_ABH_BOB}}
\end{figure*}

The final recoil velocity depends on the spin, mass ratio, and so on, and it could be as large as a few $10^3$ km/s \cite{ABH_a}. Assuming the final recoil velocity is $3000$ km/s, and the typical accelerating time is $20M$, we can get the acceleration parameter $A = 5 \times 10^{-4}$. The ringdown signal for $A=5 \times 10^{-4}$ is calculated using the method mentioned in the previous section and shown in the left panel of Fig.~\ref{GW_ABH_BOB}. We find that even for such a large kick velocity, the acceleration could not significantly influence the ringdown waveform. We are also interested in when the acceleration parameter $A$ would significantly affect GW waveforms. While the acceleration parameter $A$ is as large as $10^{-3}$, one can see the visible difference that is shown in Fig.~\ref{GW_ABH_BOB}(b). However, even in this case, the overlap is still larger than 0.99. When the acceleration reaches $10^{-2}$, it may produce a detectable effect in GWs. However, it is not clear now if some mechanisms can induce such large acceleration.

%%%%%%%%%%%%%%%%%%%%%%%%%%%%%%%%%%%%%%%%%%%%%%%%%
\section{Conclusion}\label{Conclusion}
The black hole predicted by GR has only three ``hairs": mass, spin, and charge (the well-known ``no-hair" theorem). However, alternative gravitational theories can predict more complicated (hairy) black holes. Tests of the no-hair theorem of the black hole are the critical target for GW astronomy. Developing waveform models for no-GR black holes is a foundation for testing black holes with gravitational waves. As an alternative method, we propose the PSI ($\Psi$) model based on the photon sphere to construct the waveforms, particularly for the ring-down signal in the case of a Kerr black hole. We also extend this work to more generic black holes.

Earlier in this paper, we studied the photon's motion under the Kerr metric for two cases: the equatorial and 3D planes. In the case of the equatorial plane, since $\dot{r}$ and $\dot{\theta}$ are equal to zero, the motion equations with four terms reduce to two terms, and we can get the motion orbits easily. In the case of the 3D planes, we should use two new parameters($\tilde{L}$, $\tilde{Q}$) to replace the three parameters ($E$, $L_z$, $\mathcal{Q}$), then replace $\theta$ with $\chi$ and with these new parameters we can get the motion orbits directly.

With the photon sphere at hand, we got the QNMs which are mainly described by the frequency (real part $\omega_R$) and the decay rate (imaginary part $\omega_I$). From Eq.~(\ref{omega_R_equ}), we can get the value of $\omega_R$, as shown in this equation, the value of $\omega_R$ depends on two parts: the $\theta$motion and the $\phi$motion. When calculating the value of $\omega_I$, we only need the term of $r$ direction in motion equations. We compare our results with the values of $\omega_R$ and $\omega_I$ calculated from black hole perturbation theory and find that the differences are slight for the high spin $a$. Even in the small spin case, the difference between our results and black hole perturbation is not larger than $10\%$. Because this method gives the values of $\omega_R$ and $\omega_I$ only through photon motion, we can easily extend to more generic black holes.

From $\omega_{R,~I}$, we construct the ringdown signals directly. Then we connect this part with the PN inspiral waveforms around the photon sphere and get the full inspiral-merger-ringdown (IMR) waveforms. In this way, we call this the PSI ($\Psi$) model. Since there are inspiral formalisms(e.g., SEOBNR) for Kerr black holes, we easily construct the GW waveforms from two spinning black hole mergers through the PSI model. We compare the $\Psi$ waveforms with SEOBNRv4 and SXS(NR) as shown in Figs.~\ref{compare1}, \ref{compare2}. All the overlaps are larger than 98\%, which validates that the PSI waveform model has enough accuracy. Because the method based on PSs does not need to solve the complicated black hole perturbation equations, it can be directly used on more generic black holes. For example, we first use the PS method in the KRZ metric, which describes the general axisymmetric black holes. We compare the ringdown waveforms of the KRZ black holes and the Kerr black hole (a special case of the KRZ black hole with all deviation parameters equal to zero). We find that, from the ringdown signals, we have the chance to constrain or extract the deviation parameters if they are large enough. In the next work, we will construct the full IMR waveforms for the generally axisymmetric black holes.

Finally, we use the PS to study the ringdown of accelerating black holes. The acceleration can happen in the gravitational recoil during the final merger of two black holes. We find that for the extreme recoil case, the acceleration during the kick cannot produce a detectable effect for GWs.

Connect the inspiral waveforms around the photon sphere with the ringdown part at hand. We can construct the full IMR waveforms for more generic black holes. We will apply our PSI model to no-GR black holes to get waveform templates in future work. It should be helpful in testing black holes and GR in the era of GW astronomy.

\section*{Acknowledgements}
This work is supported by The National Key R\&D Program
of China (Grant No. 2021YFC2203002), NSFC (National Natural Science Foundation of China) Grants No. 11773059 and No. 12173071. W. H. is supported by CAS Project for Young Scientists in Basic Research YSBR-006.

We would like to express my gratitude to Im\`{e}le Belahcene who helped us during the writing of this paper.

\bibliographystyle{apsrev4-1}  %% BibTeX style

%\bibliography{gravreferences}
%merlin.mbs apsrev4-1.bst 2010-07-25 4.21a (PWD, AO, DPC) hacked
%Control: key (0)
%Control: author (72) initials jnrlst
%Control: editor formatted (1) identically to author
%Control: production of article title (-1) disabled
%Control: page (0) single
%Control: year (1) truncated
%Control: production of eprint (0) enabled
%

\end{document}